\documentclass[%
reprint,
superscriptaddress,
nofootinbib,
 amsmath,amssymb,
 aps,
prc
]{revtex4-1}
\usepackage{graphicx}
\graphicspath{{./Images/}}
\usepackage{dcolumn}
\usepackage{bm}
\usepackage{slashed}
\usepackage{physics}
\usepackage{xfrac}
\usepackage{float}
\usepackage{color,xcolor}
\usepackage{mathrsfs}
\usepackage{multirow}
\usepackage[colorlinks=true,linkcolor=blue]{hyperref}%
\usepackage{hyperref}

\def\p{\bm{p}}

\definecolor{mycolor}{RGB}{0, 120, 50}

\allowdisplaybreaks[1]

\begin{document}

\preprint{APS/123-QED}

\title{Quenching jets increases their flavor}

\author{Chathuranga Sirimanna}
\affiliation{Department of Physics and Astronomy, Wayne State University, Detroit, MI 48201.}

\author{Ismail Soudi}
\affiliation{Department of Physics and Astronomy, Wayne State University, Detroit, MI 48201.}

\author{Gojko Vujanovic}
\affiliation{Department of Physics and Astronomy, Wayne State University, Detroit, MI 48201.}
\affiliation{Department of Physics, University of Regina, Regina, SK S4S 0A2, Canada}

\author{Wen-Jing Xing} 
\affiliation{Institute of Frontier and Interdisciplinary Science, Shandong University, Qingdao, Shandong, 266237, China}

\author{Shanshan Cao}
\affiliation{Institute of Frontier and Interdisciplinary Science, Shandong University, Qingdao, Shandong, 266237, China}

\author{Abhijit~Majumder} 
\affiliation{Department of Physics and Astronomy, Wayne State University, Detroit, MI 48201.}

\begin{abstract}
The widespread notion that jets quenched in a Quark-Gluon-Plasma (QGP) are similar in their parton flavor composition to jets in vacuum is critically examined. We demonstrate that while the soft to semi-hard [low to intermediate transverse momentum ($p_T$)] sector of vacuum jets are predominantly bosonic i.e., composed of gluons, \emph{sufficiently} quenched jets can have an intermediate momentum sector that is predominantly fermionic, dominated by quarks and antiquarks. We demonstrate, using leading order perturbative QCD processes, that the rate of flavor conversion from a gluon traversing the QGP as part of a jet, to a quark or antiquark, versus the reverse process, grows steadily with falling $p_T$. Simple diagrammatic estimates are followed by a variety of realistic simulations in static media, which demonstrate qualitatively similar yet quantitatively different fermion enhancements. The relation of this increase in flavor to the observed baryon enhancement at intermediate $p_T$ is studied in a fully realistic simulation. 
\end{abstract}

\date{\today}

\maketitle
\section{Introduction}
Jet quenching or the modification of hard QCD jets in a dense quark-gluon plasma (QGP)~\cite{Majumder:2010qh,Cao:2020wlm}, has reached a state of precision exploration~\cite{Kumar:2019bvr, JETSCAPE:2021ehl, JETSCAPE:2022jer}. The basic process of medium-induced energy loss leading to an increase in the number of gluon emissions from the original hard parton~\cite{Wang:1992qdg,Gyulassy:1993hr,Baier:1994bd,Baier:1996sk,  Zakharov:1996fv, Zakharov:1997uu}, was initially established by successful comparisons with data on the suppression of the scaled yield of leading hadrons ($R_{AA}$)~\cite{Bass:2008rv}, via four different  formalisms~\cite{Gyulassy:2000er,Salgado:2002cd,Arnold:2002ja, Wang:2001ifa, Majumder:2009ge}. Over time, equivalences between these various approaches were established~\cite{Majumder:2007iu,Majumder:2010qh}, and the multi-stage picture of jet modification, coupled with the concept of coherence, was developed~\cite{Cao:2017zih,Caucal:2018dla,Cao:2021rpv, Armesto:2011ir, Kumar:2019uvu}. In tandem with these developments, there arose extensive event generators based on these separate formalisms~\cite{Zapp:2008gi, He:2015pra, Majumder:2013re, Schenke:2009gb}. These event generators have also been incorporated in an end-to-end multi-stage model-agnostic event generator framework that successfully describes a majority of the available data on jet modification~\cite{JETSCAPE:2022jer, JETSCAPE:2023hqn}.

The multitude of approaches to jet modification contained within them varying descriptions of the medium. Within the last few years, all these different models of the medium have either been replaced by, or have been related to, a few transport coefficients. Focusing only on light flavors, the leading transport coefficient that encapsulates a dominant portion of the effect, the medium induces on the jet, is the transverse broadening coefficient $\hat{q}$~\cite{Baier:2002tc,Majumder:2012sh,Benzke:2012sz}, defined as the mean square momentum per unit length, exchanged between a hard parton and the medium, transverse to the direction of the hard parton:
\begin{eqnarray}
\hat{q} =  
\sum_i^N\frac{\left| \vec{k}_\perp^{i} \right|^2 }{L} . \label{eq:qhat}
\end{eqnarray}
In the equation above, a parton undergoes $N$ scatterings in a length $L$, within the QGP, without emission, with the $i^{\rm th}$ scattering imparting a transverse momentum $\vec{k}_\perp^i$.

In all formalisms of pQCD based energy loss~\cite{Gyulassy:2000er,Salgado:2002cd,Arnold:2002ja,Wang:2001ifa,Majumder:2007iu,Bass:2008rv}, the basic picture involves an increase in the amount of emissions from the hard parton(s), induced by scattering from the medium, quantified by $\hat{q}$. The flavor of the emissions, as well as the flavor of the majority of partons associated with a jet, are not {\it a priori} expected to be very different from that in the vacuum.

Most of the emissions from a hard parton in vacuum are gluons and the expectation is that, in a medium, jet partons simply radiate more gluons (i.e., partons associated with jet showers are predominantly bosonic). The goal of this paper is to demonstrate that in the plasmas typically created at RHIC and LHC, a large fraction of the emissions from jets may actually turn into quarks and antiquarks, i.e., partons associated with quenched jets may be predominantly fermionic. This is caused by the repeated re-interaction of these partons with the medium.

A jet radiates partons of all energies, starting from a fraction of the jet's own energy down to vanishingly small energies. A large fraction of the emissions are sufficiently soft that a pQCD based description of the scattering of these partons, off constituents in the medium, may not be accurate (the running of the QCD coupling~\cite{Gross:1973id,Politzer:1973fx} insists that soft partons interact strongly with the medium). A variety of strong coupling methods~\cite{Liu:2006ug, CasalderreySolana:2007qw, Chesler:2008uy, Casalderrey-Solana:2014bpa} based on the AdS/CFT conjecture~\cite{Maldacena:1997re} are currently available to describe the energy loss of a parton in a strongly interacting medium. However, in none of these approaches is the flavor of the parton affected. Thus, full jet simulations based on strong coupling approaches \emph{also} do not change the flavor profile of the jet. 

There is a trivial and somewhat obvious change in the flavor profile of the softest portion of the jet as soft partons are thermalized and lead to excitations of the medium~\cite{Tachibana:2020mtb}. Whether this thermalization is treated schematically or using strong coupling methods, one obtains the flavor composition of this portion of the jet to be similar to that of the medium itself. If the medium is fully chemically equilibrated, then these partons will show a similar quark-to-gluon ratio as in the bulk medium. 

In this paper, we focus on the intermediate energy region where the energies ($E$) or transverse momenta ($p_T$) of partons associated with the high-energy (jet) parent parton, are just above those of the bulk medium; i.e., we consider semi-hard partons, within the collection of partons, where the interaction with the medium could still be reliably calculated using pQCD (it may be the case that only part of the interaction is calculable in pQCD). Quite surprisingly, we find this region to be fermion dominated (after the jet has traversed a typical distance of about $6$~fm/$c$ in a medium held at temperature $T\sim 0.25$~GeV). In some cases, the enhancement can exceed the fermion fraction in the bulk medium itself. While this flavor conversion does not change the energy-momentum of the jet, it turns out to be a dramatic change within the particle composition of the jet: The fermion fraction within the jet can change by an order of magnitude from what it is in vacuum as will be explored. 

This paper is organized as follows: In Sec.~\ref{survey}, we recall the somewhat disconnected prior work on the topic of flavor conversion. Sec.~\ref{sec:rates} will provide a brief review of the rates of various fermion-boson conversion (or flavor conversion) processes for 2-to-2 parton scattering in pQCD at finite temperature. These rates will demonstrate that, in a thermal QCD medium, an external gluon with $E \gtrsim 8 T$  can have most of its energy transferred into a quark/antiquark at a much faster rate than a quark (or antiquark), with similar energy, can transfer most of its energy to a gluon. The choice of an ``intermediate energy" parton with $E\gtrsim 8 T$, segregates well the jet-like parton from bulk medium (thermal) partons.

The rate at which gluons radiated from a jet are converted into quarks (and antiquarks) also depends on the state of chemical equilibrium of the medium. In Sec.~\ref{sec:brick}, we carry out simulations in a static medium and produce estimates of the time when the fermion (quark + antiquark) number begins to exceed the boson (gluon) number. Given the momentum and angular structure of partons emitted by a quenched jet, the fermion excess at intermediate $p_T$ first appears as an annular ring between 0.2-1 radian away from the jet axis, after about 2-6~fm$/c$, in all models that we used (with a fermion excess appearing at lower $p_T$ in some models, at a later time). 

In Sec.~\ref{sec:realistic_simulation}, we present a variety of realistic simulations to pin down the range of angles and times where these charge/baryonic (anti-charge/anti-baryonic) rings appear. In Sec.~\ref{baryon_enhancement}, we present other measurable consequences of these charge rings, e.g., on the baryon (and anti-baryon) enhancement observed at intermediate $p_T$. A summary of the outstanding challenges to the experimental detection of these baryonic/charge rings and an outlook to future work is presented in Sec.~\ref{sec:summary_outlook}.

\section{Survey of Prior Efforts and basic formalism}
\label{survey}

Hard or semi-hard partons traversing a QGP undergo multiple scattering off constituents in the plasma. The notion that some of these interactions may lead to a flavor change (i.e. conversion) of the (semi-)hard parton has been discussed several times in the literature. In this section, we highlight these somewhat disconnected efforts and discuss the non-perturbative matrix elements that lead to these flavor conversions. While the subsequent sections will use perturbative evaluations of these matrix elements, in order to make semi-realistic estimates, we remind the reader that, similar to $\hat{q}$ [Eq.~\eqref{eq:qhat}], these conversion processes may indeed contain considerable non-perturbative contributions. 

\begin{figure}
    \centering
    \includegraphics[width=0.4\textwidth]{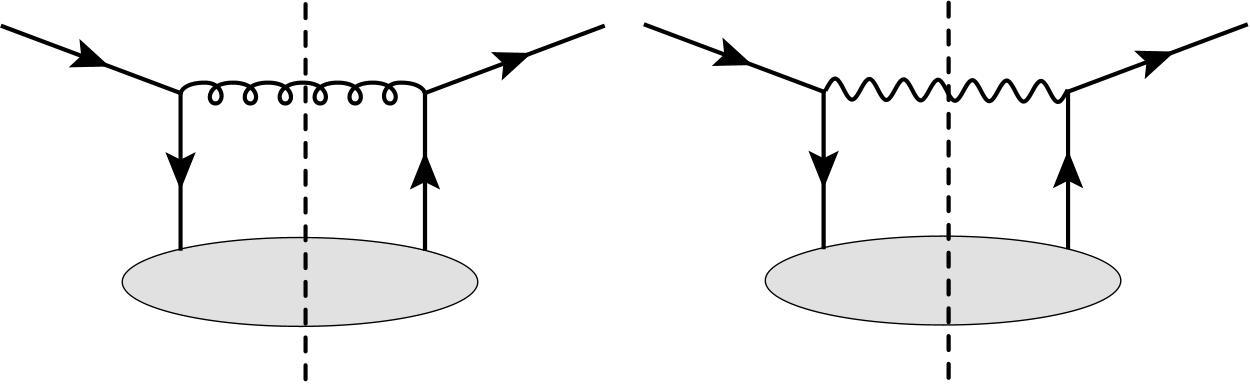}
    \caption{Non-perturbative matrix elements that lead to a semi-hard quark annihilating off a resolved soft antiquark in the medium and producing an onshell gluon (left) or photon (right).}
    \label{fig:quark_to_gluon_or_photon}
\end{figure}

The earliest examples of flavor conversion leading to formation of more quarks and antiquarks appeared in the process of strangeness enhancement~\cite{Rafelski:1982pu}. The first connection to jets was the suggestion that hard jet partons could convert to photons~\cite{Fries:2002kt} on passage through the QGP. Such photons were expected to produce a negative azimuthal anisotropy in semi-central collisions~\cite{Turbide:2005bz}. While these matrix elements were electromagnetic, the basic structure is similar to those that lead to flavor conversion. Consider the two diagrams in Fig.~\ref{fig:quark_to_gluon_or_photon}, where a semi-hard quark scatters off an antiquark in the medium producing a photon (right) or a gluon (left). The dashed line in the middle is a cut line and thus these diagrams respresent the product of an amplitude and its complex conjugate. The photon producing diagram corresponds to the work of Refs.~\cite{Fries:2002kt,Turbide:2005bz}, while the gluon producing diagrams will be considered in this paper. In either case, the soft matrix elements that control either conversion process are identical. 

\begin{figure}
    \centering
    \includegraphics[width=0.4\textwidth]{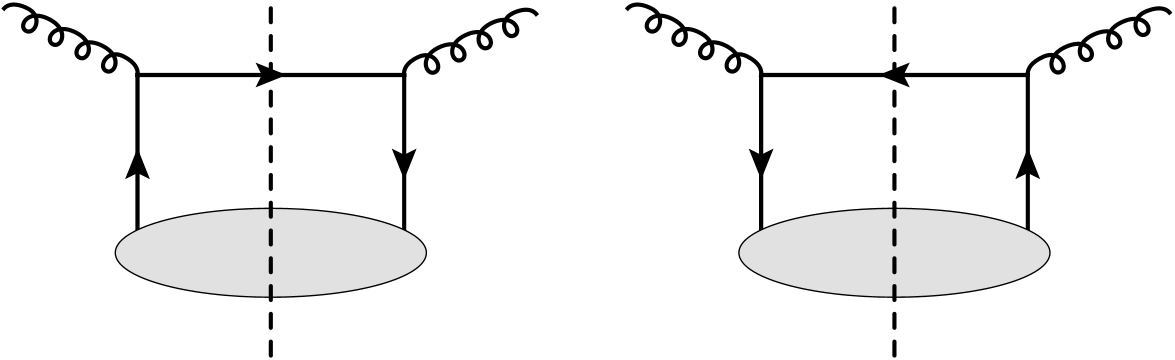}
    \caption{Non-perturbative matrix elements that lead to a semi-hard gluon converting to a semi-hard quark (left) or antiquark (right). }
    \label{fig:gluon_to_quark}
\end{figure}

The possibility of quarks converting into gluons via diagrams such as the left diagram in Fig.~\ref{fig:quark_to_gluon_or_photon} was briefly considered by Wang and Guo in Ref.~\cite{Wang:2001ifa} and later in greater detail by Schaefer et al. in Ref.~\cite{Schafer:2007xh}. Both of these efforts were in cold nuclear matter, where it was found that such contributions could be comparable to $\hat{q}$ if the quark distribution in the medium was large. These diagrams, however, only consider the conversion of a quark into a gluon. In this paper, we will point out that a much larger effect is the conversion of a gluon into a quark (or antiquark) via the diagrams in Fig.~\ref{fig:gluon_to_quark}.

The left diagram in Fig.~\ref{fig:gluon_to_quark} represents a gluon turning into a quark, while it turns into an antiquark on the right. A cursory examination of these diagrams would indicate that the soft matrix element in all the diagrams of Figs.~\ref{fig:quark_to_gluon_or_photon} and~\ref{fig:gluon_to_quark} is essentially the same. It is the expectation to find a quark or antiquark in the medium. However, the processes in Figs.~\ref{fig:quark_to_gluon_or_photon} and~\ref{fig:gluon_to_quark} have very different flavor and color degeneracies: For the case of the outgoing photon, the antiquark from the medium has to have the same color and flavor of the incoming semi-hard quark. For the case of the outgoing gluon, the antiquark from the medium has to have the same flavor as the semi-hard quark. However, for the case of the incoming semi-hard gluon, there are no flavor or color restrictions on the incoming quark or antiquark. It will thus come as no surprise that this diagram will have the largest effect in the subsequent sections.

\begin{figure}
    \centering
    \includegraphics[width=0.4\textwidth]{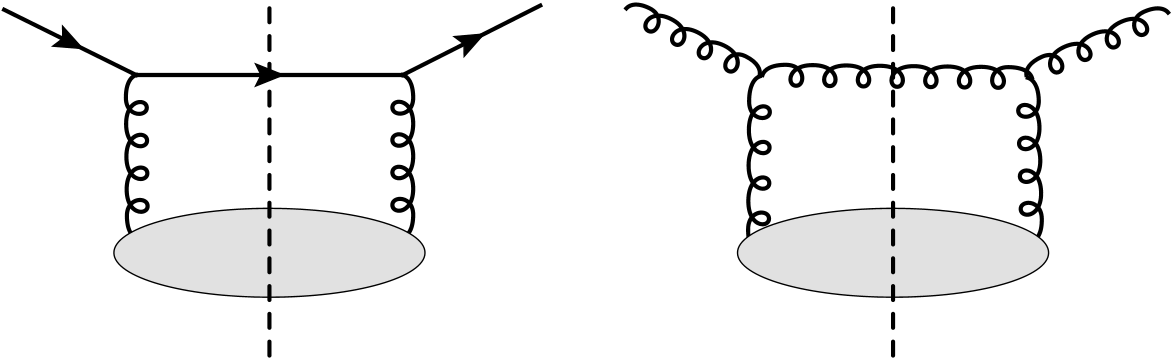}
    \caption{Gluon or Boson exchange diagrams that do not lead to flavor conversion, and are the dominant contribution to the transverse broadening coefficient $\hat{q}$ and the longitudinal coefficients $\hat{e}, \hat{e}_2$.}
    \label{fig:qhat_diagrams}
\end{figure}

Thus, all flavor changing diagrams involve the semi-hard parton exchanging a quark (antiquark) with the medium. This is in contrast with the diagrams that typically generate the transverse broadening coefficient $\hat{q}$~\cite{Baier:2002tc, Majumder:2012sh, Benzke:2012sz, Kumar:2020wvb} or longitudinal coefficients $\hat{e}, \hat{e}_2$~\cite{Majumder:2008zg}, which involve gluon exchange as shown in Fig.~\ref{fig:qhat_diagrams}. While the flavor changing diagrams do lead to transverse broadening, their main effect is the rotation of flavor of the semi-hard parton. Similar to $\hat{q}$, we can define a new transport coefficient which is sensitive to quark (antiquark) number in the QGP: The rate of flavor rotation from a quark (antiquark) to a gluon, and vice versa, can be straightforwardly derived as~\cite{Kumar:2020wvb}, 
\begin{align}
\Gamma_{q \rightarrow g } &= c_{q \rightarrow g} \frac{1}{2E_q} \int \frac{dy^- d^2 y_\perp}{(2\pi)^2}  d^2k_\perp e^{ -i \frac{k_\perp^2}{2q^-} y^-  + ik_\perp \cdot y_\perp} \nonumber \\
&\times \sum_n \frac{e^{-\beta E_n} }{Z} \langle n |  \bar{\psi}(0) \gamma^+ \psi(y^-,y_\perp) | n \rangle,  
\label{eq:quark_to_gluon} \\
\Gamma_{g \rightarrow q } &= c_{g \rightarrow q} \frac{1}{2E_g} \int \frac{dy^- d^2 y_\perp}{(2\pi)^2}  d^2k_\perp 
e^{ -i \frac{k_\perp^2}{2q^-}y^-  + ik_\perp \cdot y_\perp} \nonumber \\
&\times \sum_n \frac{e^{-\beta E_n} }{Z} \langle n |  \bar{\psi}(0) \gamma^+ \psi(y^-,y_\perp) | n \rangle. 
\label{eq:gluon_to_quark}
\end{align}
The operator expressions above can be written in gauge invariant form using a combination of light-like and transverse Wilson lines. The methodology for incorporating these lines is well known~\cite{Idilbi:2008vm,Benzke:2012sz,Majumder:2014vpa}. 
One immediately notes that both flavor conversion rates can be expressed in terms of one fundamental coefficient, i.e., 
$\Gamma_{q \rightarrow g} = c_{q \rightarrow g} \hat{f}$, and $\Gamma_{g \rightarrow q} = c_{g \rightarrow q} \hat{f}$. 
The coefficients $c_{q \rightarrow g},c_{g \rightarrow q}$ include the overall spin, and color factors related to whether the incoming (out-going) state is a quark (gluon) or gluon (quark) respectively. 
Given this structure, one immediately obtains the straightforward relation that the ratio of rates of a gluon converting to a quark (antiquark) to that of the reverse process is merely a ratio of spin, color and statistical factors (Bose or Fermi, represented as $S$):
\begin{align}
    \frac{\Gamma_{g \rightarrow q /(\bar{q} )} }{ \Gamma_{q /(\bar{q}) \rightarrow g} } &= \frac{  c_{g \rightarrow q / (\bar{q})}  S_{g \rightarrow q} }{ c_{q / (\bar{q}) \rightarrow g}  S_{q \rightarrow g}}. \label{eq:gamma_to_gamma}
\end{align}
If one ignores the statistical factors the ratio becomes a pure number.

The ratio in Eq.~\eqref{eq:gamma_to_gamma} will be evaluated several times in the subsequent sections with different restrictions on the included processes, both with and without statistical factors, in static and dynamic media. We are not the first to evaluate these or point out the possibility of flavor conversions of jets traversing a QGP. These were first discussed in a series of papers by Liu et al.~\cite{Ko:2007zz,Liu:2007aj,Liu:2008zb}. These authors studied the possibility that the leading parton in a jet could convert from quark to gluon or vice versa. While the rates were not small, they tended to drop with increasing $p_T$ or energy of the parton. 

As will be shown in this paper, the region where the effect of conversions is most dramatic is the semi-hard region with momenta just above $8T$, where $T$ is the temperature of the medium. Some of our approach is similar to the work of Refs.~\cite{Ko:2007zz,Liu:2007aj,Liu:2008zb}, however we will focus on the shower of gluons radiated by the hard parton, and not the hard parton itself. In our case, one would trigger on a hard parton and observe the change in the flavor of the radiated gluons as they propagate through the medium. 

The flavor or chemical change of full distributions of partons correlated with a hard parton traversing and equilibrating within a dense medium have been extensively studied by one of us in collaboration with Schlichting and Mehtar-Tani~\cite{Schlichting:2020lef, Mehtar-Tani:2022zwf}, using the finite temperature rates derived by Arnold, Moore and Yaffe~\cite{Arnold:2000dr,  Arnold:2002zm}. However, these calculations focused on partons with asymptotic energies $\gtrsim 500 T$ and thus observed a much delayed onset of the quark antiquark numbers becoming comparable to the gluon numbers. 

In the subsequent sections, we will predominantly focus on partons with energy ranging from $2~{\rm GeV} \lesssim E \lesssim 5$~GeV radiated from a jet with $E\gtrsim 25$~GeV. For most of the simulations, the medium will be static with $T= 0.25$~GeV. Thus, scaling with the temperature $T$, the jet has an energy of $100T$ and we are focusing on partons radiated from the jet with energy above $8T$ and less than $20T$. Partons with momenta above $20T \simeq 5$~GeV will be considered hard in this effort. Partons with $E \leq 8 T \simeq 2$~GeV will be considered as soft. While pQCD based estimates on their population are presented below, these are done only for completeness. It will be assumed that partons with $E\leq 8T \simeq 2$~GeV will eventually thermalize and hadronize as a part of the bulk medium. Our main focus will remain on the semihard region with $8T \lesssim E \lesssim 20T$ (or $2~{\rm GeV} \lesssim E \lesssim 5$~GeV for $T=0.25$~GeV).

Path lengths in the medium range from 2 to 10 fm/$c$. The path lengths and temperatures are representative of the average temperature and path lengths experienced by jets at RHIC and LHC. The reason to not take the energy of the jet to be too large is to reduce the amount of vacuum emission that the partons escaping the medium will produce. Thus, picking a jet with an $E\sim 100T$ in a medium of length around 2-10~fm/$c$ produces a prominent enhancement in the quark and antiquark number, correlated with the jet.

While the rates in Eqns.~(\ref{eq:quark_to_gluon},\ref{eq:gluon_to_quark}) are cast in terms of non-perturbative matrix elements, in the remainder of this paper, we will evaluate these using perturbation theory. There is every indication that for partons with energies in the region $E \gtrsim 2$~GeV, the interaction with the medium may predominantly be non-perturbative~\cite{Zhao:2021vmu,Fries:2003vb,Hwa:2002tu,Molnar:2003ff}. The calculations in this paper should thus be considered as an estimate of the effect of these terms. We will show that the flavor conversion processes will produce more than an order of magnitude increase in the number of quarks and antiquarks that are correlated with the jet (compared to a jet in vacuum). Thus, while our perturbative estimates may not be completely accurate, the magnitude of the effect indicates that this is a large effect which will survive modification of the calculation scheme. 

In this first effort to understand the enhancement of fermion number within a jet shower, we will not attempt to hadronize the simulated jets. The change in jet hadrochemistry as a signal for jet quenching has been pointed out before~\cite{Sapeta:2007ad}, though not for the reasons that will be highlighted in this paper. Most of the results in this paper will be partonic. The increase in the number of quarks and antiquarks within the jet shower will make hadronization via the standard process of Lund string breaking~\cite{Andersson:1997xwk} unfeasible. Na\"{i}vely, given the small number of partons within the jet cone, the most obvious signal would be event-by-event fluctuations of charge and baryon number.  However, any current hadronization module will itself introduce modifications to this effect. The study of the effect of hadronization on these fluctuations will appear in a future effort.

\section{Rates of flavor exchange with the medium}
\label{sec:rates}

\begin{figure}
    \centering
    \includegraphics[width=0.45\textwidth]{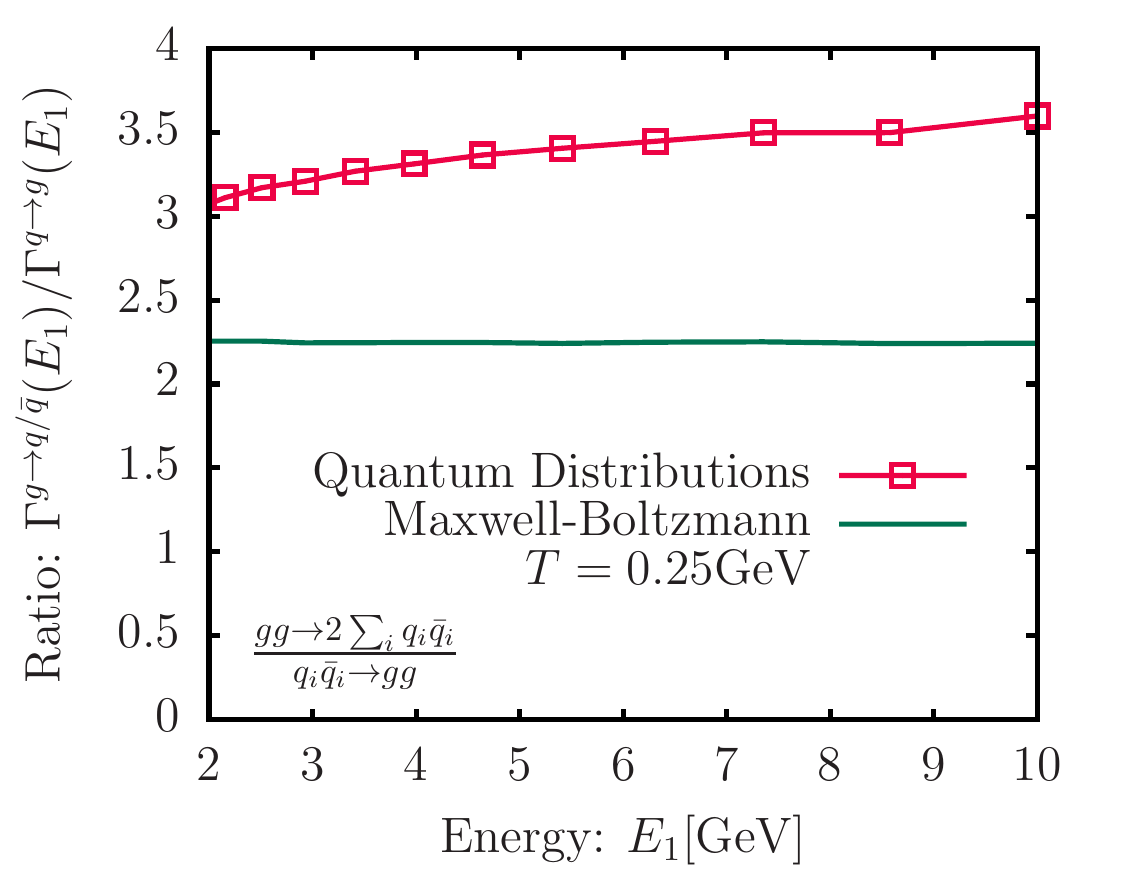}
    \includegraphics[width=0.45\textwidth]{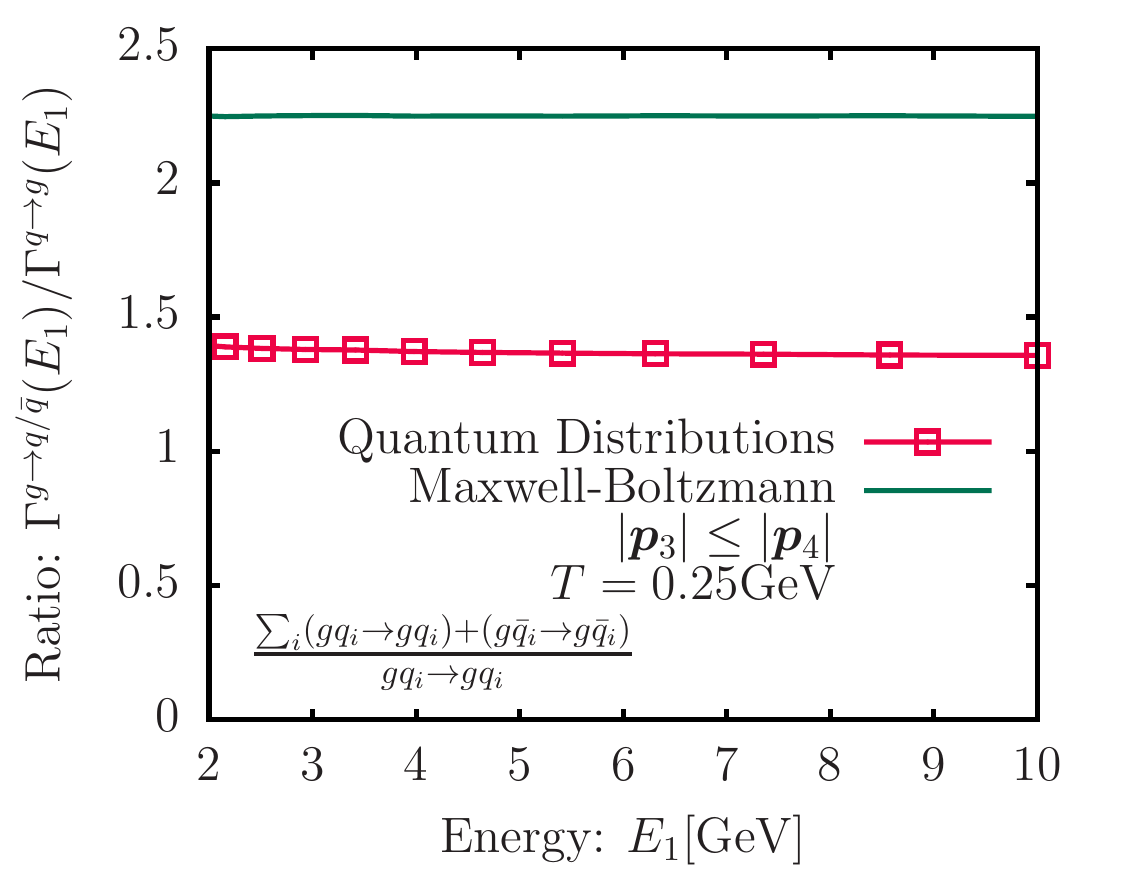}
    \caption{(Color online) Ratio of the rate of a semi-hard gluon converting to a semi-hard quark (or antiquark) to that of a semi-hard quark (or antiquark) converting to a semi-hard gluon. The top panel compares the ratio from only the diagrams that describe the process of gluon fusion to quark antiquark to that of quark antiquark annihilation into gluons Eq.~(\ref{eq:RatioGGToQQ}). The lower panel considers the ratio of quark gluon scatterings as calculated in Eq.~(\ref{eq:RatioQGToQG}). While the green lines represents the ratio where the equilibrium distributions are taken to be the classical Maxwell-Boltzmann distribution which leads to an exact ratio of $2.25$, an energy dependence is obtained for the rates computed using the equilibrium distribution from quantum statistics shown in the red lines. }
    \label{fig:Ratio}
\end{figure}

The perturbative description of multiple elastic interactions of a hard QCD parton (with energy $E\gtrsim 10T$) with a QGP medium (at temperature $T$) can be cast in terms of an effective kinetic description, with the collision term obtained from tree-level diagrams. The rate of scattering of a \emph{single} hard parton $a$, with four-momentum $p_1$, with a medium parton $b$, with four-momentum $p_2$, yielding outgoing partons $c$ and $d$ with four-momenta $p_3$ and $p_4$ can be expressed as, 
\begin{align} 
    \Gamma_{ab\to cd} =
    &\int \frac{d^3 \p_2}{(2\pi)^3}\frac{d^3 \p_3}{(2\pi)^3}\frac{d^3 \p_4}{(2\pi)^3} \label{eq:RateEq} 
    f_b(\p_2) [1\pm f_c(\p_3)]
    \\
    &\times\frac{|\mathcal{M}_{ab\to cd}|^2}{16 E_1 E_2 E_3 E_4 }
    (2\pi)^4 \delta^{(4)}\left(p_1 + p_2-p_3-p_4\right),\nonumber
\end{align}
where incoming thermal parton and outgoing partons are integrated over. In the equation above, all degeneracy (i.e. spin and color) factors are absorbed into the matrix element squared $|\mathcal{M}_{ab\to cd}|^2$, and accordingly such factors are removed from the distribution functions $f_{b,c}(\p)$. Note that the $f_{i}(\p)$ notation emphasizes that parton momenta satisfy the on-shell condition $p^2=0$ (assuming massless partons). In the case of thermal equilibrium, these are given by the Bose-Einstein distribution $f_g(\p)= n_g(\p)=\left[ e^{p\cdot u/T}-1\right]^{-1}$ for gluons, and the Fermi-Dirac distribution $f_q(\p) = \Tilde{n}_q(\p) = \left[ e^{p\cdot u/T} +1 \right]^{-1}$ for quarks and antiquarks, where $u^\mu$ is the local fluid velocity.

At leading order of the QCD coupling, the processes can be separated into two classes.
The dominant processes involve completely elastic scattering, where the boson-fermion nature of both scattering partons is unchanged: The semi-hard incoming quark (gluon) remains a semi-hard incoming quark (gluon) with a mild change in its 4-momentum. These lead to diffusion and drag of the hard parton's energy with schematic rates $\Gamma_{\rm Diffusion} \propto T^2 \partial^2_{\p_1} f_a(\p_1)$ and $ \Gamma_{\rm Drag} \propto  T^3 \partial_{\p_1} f_a(\p_1)$, respectively \cite{Schlichting:2020lef}. However, they will not cause any change to the flavor profile of the jet. We instead focus on the special case of flavor conversion, where the outgoing particle with energy comparable to the semi-hard parton is of different quantum statistics than the initial hard parton $a$. At high energies, these conversion processes are suppressed by an inverse power of momentum as $\Gamma_{\rm Conversion} \propto \frac{T^2}{|\p_1|} f_a(\p_1)$; however, for semi-hard partons this can contribute significantly.

When the semi-hard parton is a quark (antiquark), these consist of three processes: Quark-antiquark annihilation into gluons, $Q_i \,\, \bar{q}_i \rightarrow G\,\,g$ ($\bar{Q}_i \,\, q_i \rightarrow G \,\, g$), quark gluon scattering, $Q_i \,\, g \rightarrow q_i \,\, G$, and antiquark gluon scattering, $\bar{Q}_i \,\, g \rightarrow \bar{q}_i \,\, G $. Note that we use the notation where capital letters indicate the semi-hard parton and lowercase letters are reserved for soft particles. (For the remainder of this study, the identity of the hard parton should be clear from the context). When the semi-hard parton is a gluon, the reverse processes include pair production, $ G \,\, g \rightarrow Q_i \,\, \bar{q}_i   $ ($ G \,\, g \rightarrow \bar{Q}_i \,\, q_i   $), and gluon scattering with a quark, $G \,\, q_i \rightarrow Q_i \,\, g$, or antiquark, $G \,\,\bar{q}_i \rightarrow \bar{Q}_i \,\, g$. The matrix elements for all these processes are listed in Table~\ref{tab:MatrixElements}.

\begin{table}
\begin{tabular}{ c|c  }
$ab \to cd $ & $ \nu_{b}\sum_{\nu_{c} \nu_{d} }|\mathcal{M}_{ab\to cd}|^2/g^4_s$\\
 \hline
 $gg \to 2\sum_i q_i \bar{q}_i $ & $2 N_f C_F  \left( \frac{u}{t} + \frac{t}{u} - \frac{C_A}{C_F}  \frac{t^2 + u^2 }{s^2}\right) $ \\ \hline
 $q_i \bar{q}_i \to gg$ & $ 2C_F^2 \left( \frac{u}{t} + \frac{t}{u} - \frac{C_A}{C_F}  \frac{t^2 + u^2 }{s^2}\right) $ \\ \hline
 $\sum_i g q_i \to gq_i $ &  \multirow{2}*{ $-N_f C_F \left( \frac{u}{s} + \frac{s}{u} \right) + N_f C_A\frac{s^2 + u^2}{t^2} $ }  \\ 
 $\sum_i g \bar{q}_i \to g\bar{q}_i $ & \\ \hline
 $q_ig \to q_i g $ &  \multirow{2}*{$- 2C_F^2 \left( \frac{u}{s} + \frac{s}{u} \right) + 2C_FC_A \frac{s^2 + u^2}{t^2}$}  \\
 $\bar{q}_i g \to \bar{q}_i g $ & \\ 
\hline
\end{tabular}
\caption{Fermion-Boson conversion matrix elements: We average over the degrees of freedom of the initial hard parton $a$ and sum over the initial medium parton $b$ and final state partons $c$ and $d$. The conventional definition of Mandelstam variables \cite{Peskin:1995ev} is used.}
 \label{tab:MatrixElements}
 \end{table}


In the subsequent subsections, the rates of the flavor-conversion processes mentioned above where a semi-hard gluon turns into a semi-hard quark (antiquark) will be compared with the reverse processes for each, i.e., a semi-hard quark (antiquark) turning into a semi-hard gluon. The matrix elements in Table~\ref{tab:MatrixElements} will be integrated over the momenta of the incoming and outgoing soft thermal partons. Both rates will be compared within a static QCD medium held at a fixed temperature. In all cases, the rate for gluons to turn into a quark (antiquark) is found to be higher than the reverse process. 

\subsection{QCD annihilation processes} 

A hard gluon fusing  with a thermal gluon and producing a quark antiquark pair is considered first. This conversion rate is given by the double sum over quark flavors, since the final semi-hard parton can be either a quark or an antiquark. For the purposes of the numerical simulations of this process, done using the JETSCAPE (Jet Energy-loss Tomography with a Statistically and Computationally Advanced Program Envelope) framework, only the soft particle distributions are integrated over, yielding:
\begin{align} 
    \Gamma_{gg \to 2\sum_i q_i \bar{q}_i} &= 
    \int \frac{d^3 \p_2}{(2\pi)^3}\frac{d^3 \p_3}{(2\pi)^3}\frac{d^3 \p_4}{(2\pi)^3} \nonumber\\
    \times & f_g(\p_2) [1-f_q(\p_3)]\frac{|\mathcal{M}_{gg\to 2\sum_i q_i \bar{q}_i}|^2}{16 E_1 E_2 E_3 E_4 } \nonumber \\
    \times & (2\pi)^4 \delta^{(4)}\left( p_1 + p_2-p_3-p_4\right) \;. \label{eq:RateGGToQQ}
\end{align}
In the equation above, the flavor-summed matrix element square is defined as $|\mathcal{M}_{gg\to 2\sum_i q_i \bar{q}_i}|^2 = \sum_{i}^{N_f}|\mathcal{M}_{gg\to q_i\bar{q}_i}|^2 + |\mathcal{M}_{gg\to \bar{q}_i q_i}|^2$, where the subscripts keep track of the momentum of the semi-hard gluon being transferred to the quark ($g \,\, g \rightarrow q_i \,\, \bar{q}_i$) or to the antiquark ($g \,\, g \rightarrow \bar{q}_i \,\, q_i$).

In both cases above, a sum over flavors is present, regardless of whether the hard momentum transfers to the quark or the antiquark. Conversely, the annihilation of a hard quark with a medium antiquark does not involve a sum of the quark flavors, since the quark-antiquark flavor must match, thus giving 
\begin{align}
    \Gamma_{q_i \bar{q}_i \to gg } &= 
    \int \frac{d^3 \p_2}{(2\pi)^3}\frac{d^3 \p_3}{(2\pi)^3}\frac{d^3 \p_4}{(2\pi)^3} f_q(\p_2) [1+f_g(\p_3)] \label{eq:RateQQToGG} \\
    &\times\frac{|\mathcal{M}_{q_i \bar{q}_i \to gg} |^2}{16 E_1 E_2 E_3 E_4 }(2\pi)^4 \delta^{(4)}\left(p_1 + p_2-p_3-p_4\right) \;. \nonumber
\end{align}

If we neglect differences between the thermal quark and gluon distributions, as well as quantum statistics, and use Maxwell-Boltzmann (MB) distributions for both quark or gluon, we would get, 
\begin{align}
    & f_g(\p_2) [1-f_q(\p_3)]\rightarrow f_{MB}(\p_2), \nonumber \\
    &  f_q(\p_2) [1+f_g(\p_3)] \rightarrow f_{MB} (\p_2).   
\end{align}
With these substitutions, the only difference between the integrands of Eqs.~\eqref{eq:RateGGToQQ} and \eqref{eq:RateQQToGG} are the degeneracy factors of the matrix element. Thus, the ratio of a semi-hard gluon converting to a semi-hard quark (antiquark) to the reverse process of conversion of a semi-hard quark (or antiquark) into a gluon, via the matrix elements that describe annihilation or fusion, is given by
\begin{align}\label{eq:RatioGGToQQ}
    \frac{\Gamma_{gg \to 2\sum_i q_i \bar{q}_i}}{\Gamma_{q_i \bar{q}_i \to gg }} \simeq \frac{N_f}{C_F} = 2.25\;.
\end{align}
In the equation above, $C_F = 4/3$ is the quadratic Casimir in the fundamental representation, and $N_f=3$  is the number of quark flavors.

Using full quantum statistics increases the ratio by more than 50\% and also introduces a mild dependence on the energy of the semi-hard parton, as shown in the upper panel of Fig.~\ref{fig:Ratio}. Depending on the strength of the overall rate, due to the medium, the shower of a hard parton will, over time, contain  more and more semi-hard quarks (and antiquarks) even when starting with an initial hard gluon.

\subsection{QCD Compton scattering}
Compton scattering in the medium is another process that can change a semi-hard fermion into a semi-hard boson. In the current kinematic limit, whenever the momentum of the quark internal line is soft, i.e., the Mandelstam variable $u\to 0$, the conversion rate (fermion-to-boson or vice versa) is enhanced. For a semi-hard gluon scattering with a medium quark or antiquark, the rate is:
\begin{align} 
    \Gamma_{\sum_i g q_i \to  g q_i } & = 
    \int \frac{d^3 \p_2}{(2\pi)^3}\frac{d^3 \p_3}{(2\pi)^3}\frac{d^3 \p_4}{(2\pi)^3} f_q(\p_2) [1+f_g(\p_4)]\nonumber\\
    \times\,&\frac{|\mathcal{M}_{\sum_i g q_i \to  g q_i}|^2}{16 E_1 E_2 E_3 E_4 } \label{eq:RateGQToGQ} \\
\times\,& (2\pi)^4 \delta^{(4)}\left(p_1 + p_2-p_3-p_4\right) \Theta(|\p_4|-|\p_3|) \;.\nonumber
\end{align}
While for the reverse process of a semi-hard quark scattering with a medium gluon, the rate is given by 
\begin{align} 
    \Gamma_{ q_i g \to  q_i g}  &= 
    \int \frac{d^3 \p_2}{(2\pi)^3}\frac{d^3 \p_3}{(2\pi)^3}\frac{d^3 \p_4}{(2\pi)^3} f_g(\p_2) [1-f_q(\p_4)]\nonumber\\
    &\times\frac{|\mathcal{M}_{ q_i g \to  q_i g}|^2}{16 E_1 E_2 E_3 E_4 } \label{eq:RateQGToQG}\\
    &\times(2\pi)^4 \delta^{(4)}\left( p_1 + p_2-p_3-p_4\right) \Theta(|\p_4|-|\p_3|) \;. \nonumber
\end{align}

If we consider the momentum exchange to be small $(q=p_1-p_3\ll p_1,p_2)$, the dominant contributions to the quark gluon scattering are the terms proportional to $\frac{s}{u}$ and $\frac{s^2+u^2}{t^2}$ in Tab.~\ref{tab:MatrixElements}. Because the energy $(\sim |q|)$ gained by the medium parton is small compared to the energy of the semi-hard parton, only the first term will lead to flavor conversion, while the second term will typically contributes to energy loss. 
In this section, we will consider the full matrix element. Hence, to identify flavor converting processes, we will apply a kinematic selection that the out-going parton with different flavor than the initial hard parton takes a greater fraction of the energy.

Using the above two rates, the ratio of a semi-hard gluon converting to a quark (antiquark) to the reverse process is exactly the same as the ratio of rates in the preceding subsection, namely
\begin{align}
    \frac{\Gamma_{\sum_i g q_i \to  g q_i}+\Gamma_{\sum_i g \bar{q}_i \to  g \bar{q}_i}}{\Gamma_{q_i g \to q_i g }} \simeq \frac{N_f}{C_F} = 2.25\;.\label{eq:RatioQGToQG}
\end{align}
In the equation above Maxwell Boltzmann statistics were used once again. Even though using quantum statistics for the process actually reduces the ratio, it still remains above $1$, as shown in the lower panel of Fig.~\ref{fig:Ratio}.
\subsection{From a gluon shower to a quark (antiquark) shower}
The two preceding subsections clearly demonstrate that the rate for a semi-hard gluon to turn into a semi-hard quark or antiquark is much larger than for the reverse process. In this subsection, we will try to estimate the fate of the semi-hard gluons emitted from a jet as it propagates through an equilibrated QGP. 

In this first effort, we focus on jets with an energy $E_J\simeq 25$~GeV. The reason is that these jets have a large enough energy that they will definitely radiate several soft gluons on traversal through the QGP. Also, their energy is not that high that a considerable portion of the jet will continue to radiate outside the medium and produce a dominant gluon shower in the vacuum~\cite{Majumder:2013re}. 
A jet with an energy $E_J\simeq 25$~GeV~\footnote{We use $\simeq$ as the energy of the jet originating parton is set as 25~GeV, which is not equal to the energy of the clustered jet.} will radiate several gluons with energies 2~GeV$\lesssim E \lesssim 5$~GeV (See Fig.~\ref{fig:Vacuum}). These gluons will multiply scatter and interact with medium. In this process, they may convert into a quark or antiquark. The produced quarks and antiquarks may also scatter and could in turn convert back into gluons.

\begin{figure}
    \centering
    \includegraphics[width=0.45\textwidth]{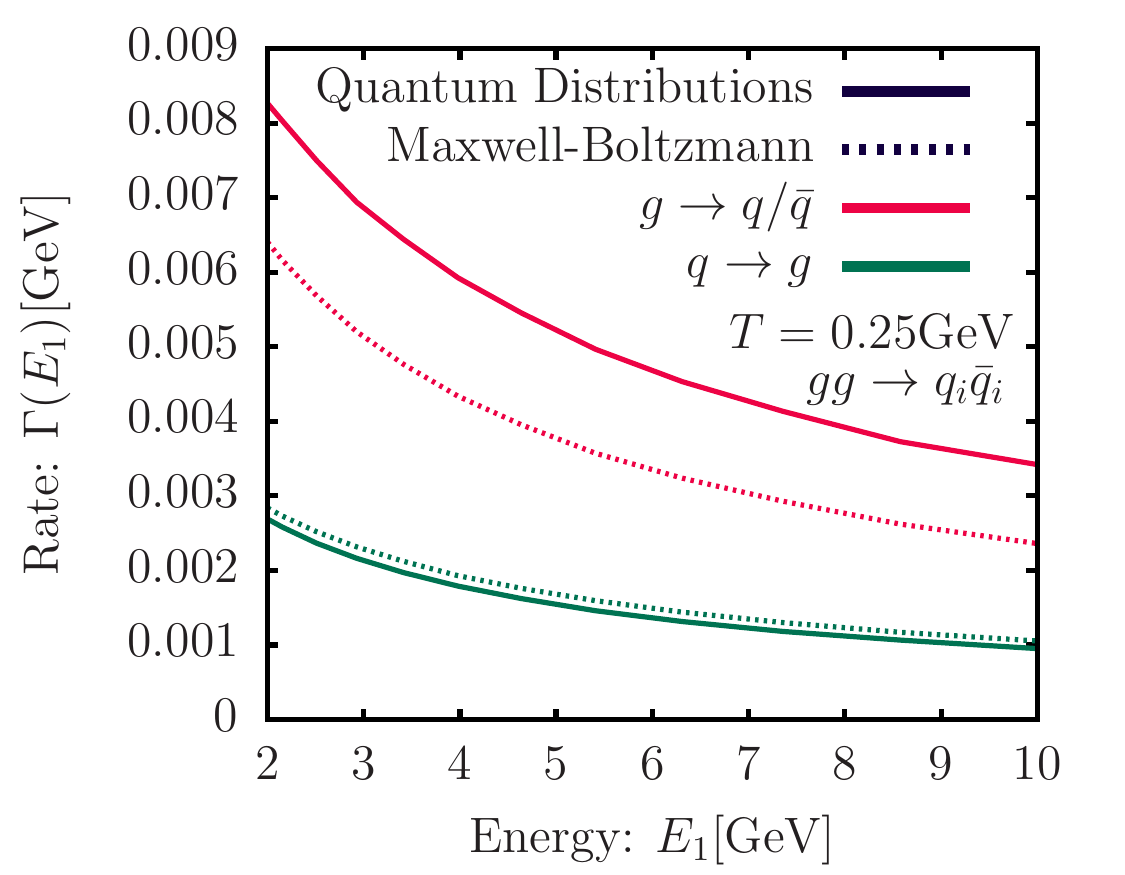}
    \caption{(Color online)
    QCD annihilation processes. The rate of gluon conversion into quark (or antiquark) from Eq.~(\ref{eq:RateQQToGG}) is represented in red lines, while the reverse process from Eq.~(\ref{eq:RateGGToQQ}) is displayed in green lines. 
    The dashed lines represent the rate where the equilibrium distributions are taken to be the classical Maxwell-Boltzmann distribution, and the rates computed using the equilibrium distribution from quantum statistics are shown in the full lines. }
    \label{fig:Rate1}
\end{figure}

\begin{figure}
    \centering
    \includegraphics[width=0.45\textwidth]{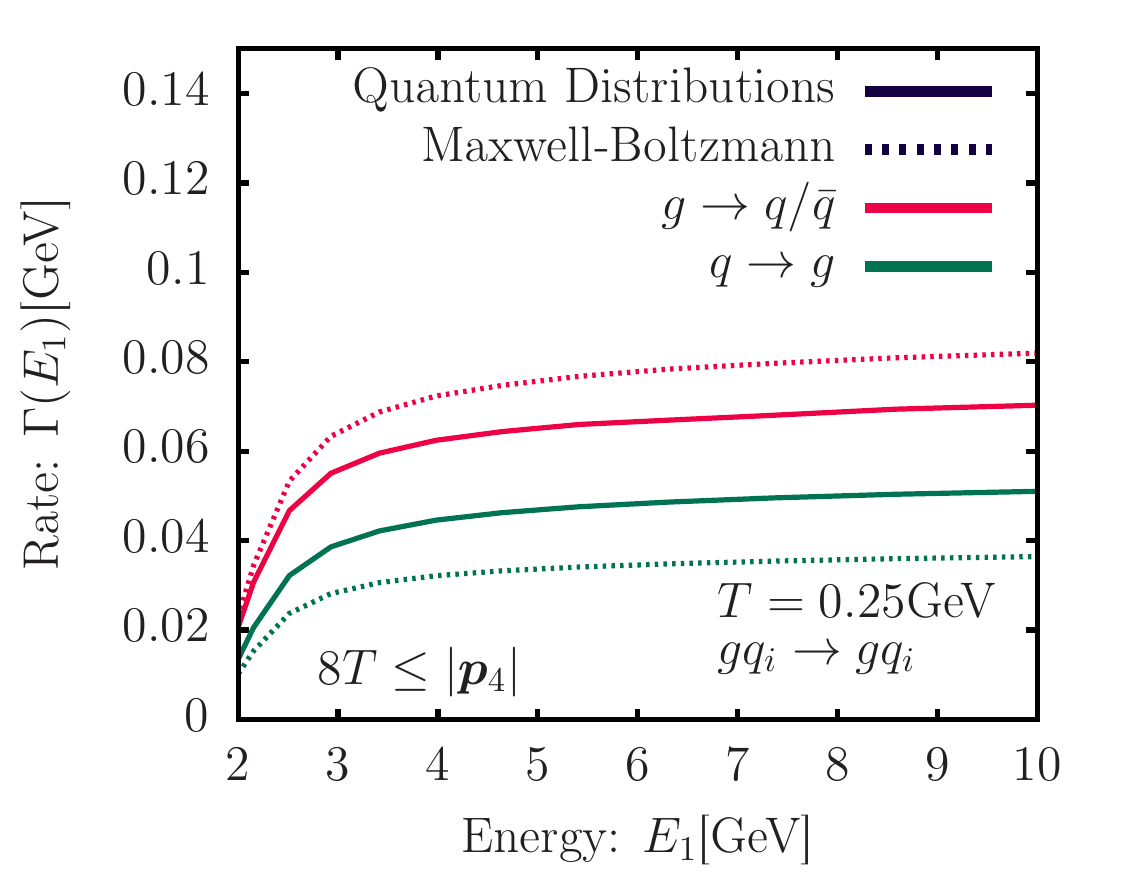}
    \includegraphics[width=0.45\textwidth]{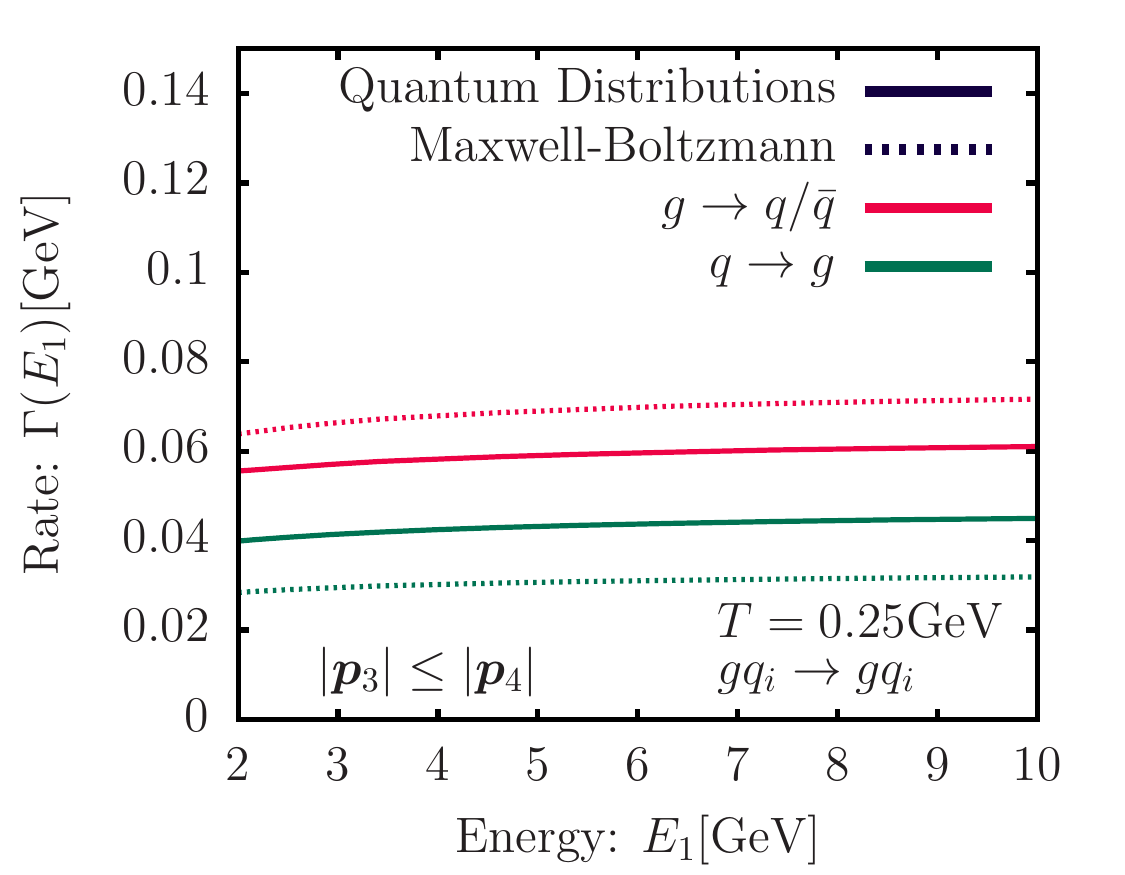}
    \caption{(Color online)
    QCD Compton scattering.
    The rate of gluon conversion into quark (or antiquark) Eq.~(\ref{eq:RateQGToQG}) is represented in red lines, while the reverse process from Eq.~(\ref{eq:RateGQToGQ}) is displayed in green lines. 
    Kinematic selections are employed for the momentum $p_4$ of the semi-hard parton with opposing flavor, in the top panel $8T\leq |\p_4|$ or $|\p_3|\leq |\p_4|$ in the bottom panel.
    While the dashed lines represents the rate where the equilibrium distributions are taken to be the classical Maxwell-Boltzmann distribution, the rates computed using the equilibrium distribution from quantum statistics are shown in the full lines. }
    \label{fig:Rate2}
\end{figure}

To estimate the probability of conversion, we go beyond the ratios of rates and plot the absolute rates, in a thermal medium, in Figs.~\ref{fig:Rate1} and~\ref{fig:Rate2}. As the case for the ratio of rates, we continue to set the temperature of the medium to be $T=0.25$~GeV. In Fig.~\ref{fig:Rate1}, we plot the rates for two gluons to pair produce a quark or antiquark (red lines), the solid line includes quantum statistics for the incoming and outgoing soft partons, while the dotted line assumes Maxwell Boltzmann statistics. The green lines represent a quark or antiquark annihilating with its antiparticle and producing two gluons. Including the effect of quantum statistics, the rate for a gluon to convert into a quark (or antiquark) can be almost 3 times as high as the reverse (quark to gluon) process in the region with 2~GeV$\lesssim E \lesssim 5$~GeV; the overall rates are rather small ($\sim 0.008$~GeV) however. Thus, these rates will act as an additive correction to the Compton process.

In Fig.~\ref{fig:Rate2}, we plot the conversion of a gluon into a quark or antiquark (red lines) and vice-versa (green lines) from the Compton process. The solid and dashed lines indicate rates with and without quantum statistics, as described above. In the Compton process, a semi-hard gluon interacting with a thermal quark could produce a semi-hard quark and a semi-hard gluon simultaneously. Hence, we present two separate rates: The top panel indicates the rate for a parton to be produced with an energy $|\p_4| > 10 T$ with a different flavor than the semi-hard projectile. Red lines indicate that the projectile is a gluon, and green lines are for a quark. The bottom panel plots the rate of the semi-hard projectile converting its flavor [gluon to (anti)quark or (anti)quark to gluon], where the outgoing semi-hard parton (or parton with larger energy) has a different flavor than the projectile. In both panels, the $x$-axis is the energy of the semi-hard projectile ($E_1$).

We conclude this section with a numerical estimate of the physical rate for a semi-hard gluon (with $2\lesssim E \lesssim 5$~GeV) to convert into a semi-hard quark (antiquark) and vice versa. Using either panel in Fig.~\ref{fig:Rate2}, we note that the rate for a semi-hard gluon to produce a semi-hard quark or antiquark (whether or not the fermion is the leading outgoing parton) via the Compton process is about 0.06~GeV (see mean of the two red lines). Combining the rate from pair creation (Fig.~\ref{fig:Rate1}), the rate for a semi-hard gluon to produce a semi-hard quark or antiquark is 
\begin{align}
    R_{g \rightarrow q+\bar{q}} \simeq 0.07 {\rm GeV} \simeq 0.35 / {\rm fm}.
\end{align}
 The rate for the reverse process is about half of this (using the green lines in either plot from Fig.~\ref{fig:Rate2} and including the rates from the plot in Fig.~\ref{fig:Rate1}). We have also used natural units $\hbar c \simeq 0.2$~GeV$\cdot$fm. 
 
 The rate in the above equation is remarkably large. It implies that a semi-hard gluon will definitely convert to a quark (or antiquark) by traversing a mere 3~fm of a QGP at $T = 0.25$~GeV. Of course, if the medium were longer, the final population would eventually tend towards twice as many quarks and antiquarks compared to gluons.

By any measure, the estimate in the preceding paragraphs presents a rather startling effect. It should completely disabuse one of the notion that a jet in a medium is a central hard parton surrounded by a gluon shower. In the subsequent sections, we will study this effect using the solution of the Boltzmann equation, followed by simulations using LBT and MATTER+MARTINI within the JETSCAPE framework. In all cases, we will note that a large portion of the gluons in a jet shower, in a medium, are converted to quarks and antiquarks. 


\section{The effective Boltzmann equation for a jet in a static QGP}
\label{sec:brick}
\begin{figure}
    \centering
    \includegraphics[width=0.4\textwidth]{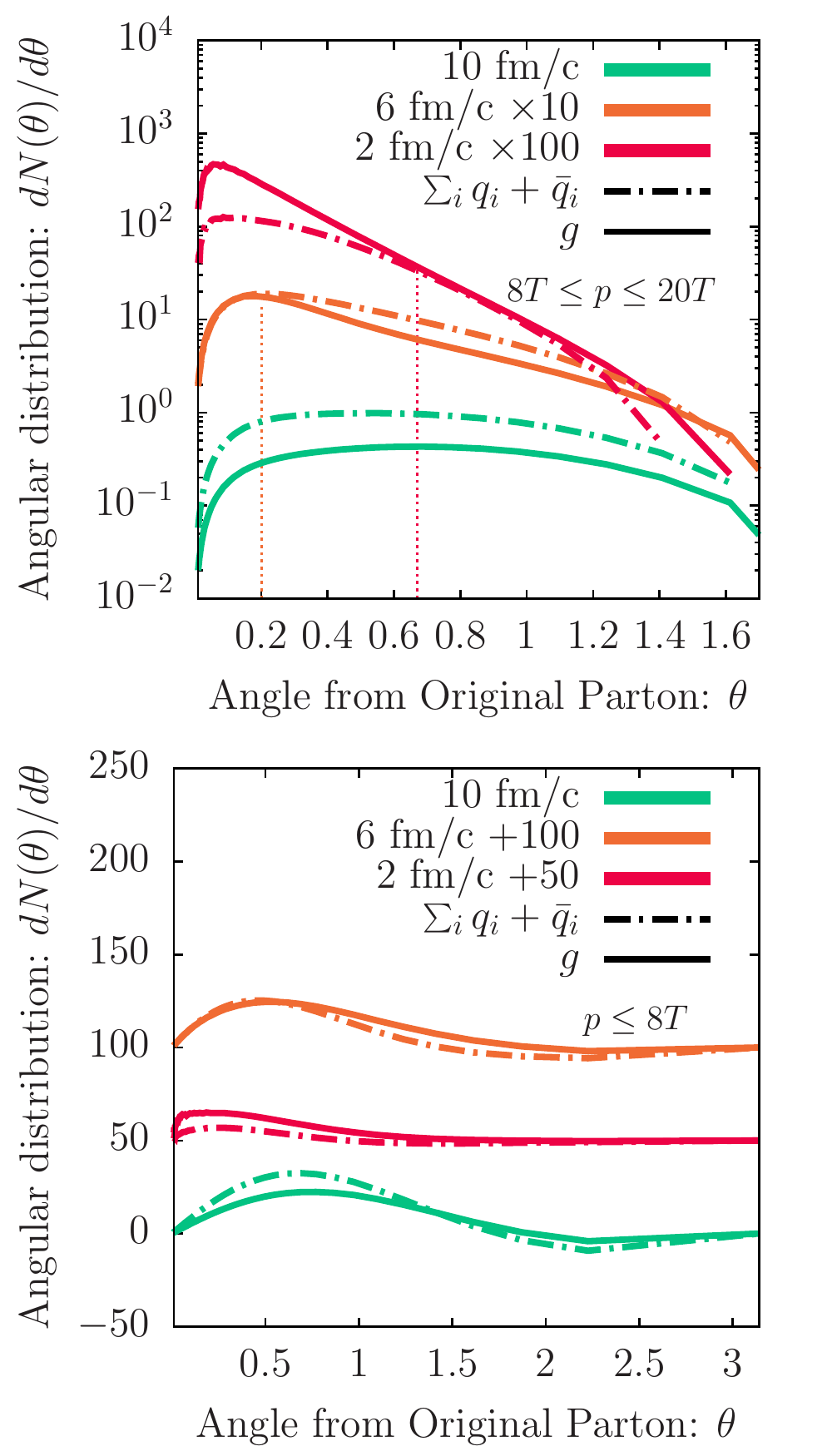}
    \caption{(Color online) Particle number distribution as a function of the angle from the original parton, for hard partons with momentum $8T\leq p\leq 20T$ (top) or soft parton with momentum $p\leq 8T$ (bottom) with $T=0.25$~GeV. The gluon distribution is displayed in full-lines, while the sum of quarks and antiquarks is displayed in dashed-lines at different times $t=2, 6$ and $10$~fm/$c$. In the top panel, the vertical dashed lines indicate the angle where the fermion distribution crosses the gluon distribution.}
    \label{fig:KineticEvol}
\end{figure}
In this section, we will investigate the evolution of hard partons in a static QGP medium using an effective kinetic description of QCD, at leading order. Based on the approach of Arnold, Moore and Yaffe (AMY) \cite{Arnold:2002zm}, we study the evolution of the phase-space distribution $f(\p)$, where, after integrating out position space, the kinetic equation is given by 
\begin{align}
    \partial_t f_a(\p) = C_a^{2\leftrightarrow 2}[f]+ C_a^{1\leftrightarrow 2}[f]\;.
\end{align}
The leading-order QCD elastic scatterings are described by the $2 \leftrightarrow 2 $ collision integral $C_a^{2\leftrightarrow 2}[f]$, where we use hard thermal loop propagators for the internal quark and gluon to regulate the divergent small angle scatterings, while the other in-coming and out-going parton lines are assumed to be vacuum like~\cite{Arnold:2000dr}, with thermal distributions for the partons that emerge from and re-enter the medium.

Multiple scattering of a hard parton with the medium can cause the parton to become slightly off-shell. The hard parton loses its off-shellness via radiation, which is enhanced in the collinear region. These infinite number of diagrams, iteratively including an arbitrary number of scatterings, can be resummed into an effective $1\leftrightarrow 2$ radiation / absorption rate, which is described by the collision integral $C_a^{1\leftrightarrow 2}[f]$.
This medium-induced radiation is governed by an interplay between the medium scale given by the mean free path $\lambda_{\rm mfp} \sim 1/m_D$ and the formation time of the radiation $t_f \sim \sqrt{2zE_{\p}/\hat{q}}$, which leads to a time-dependent rate of radiation in the collision integral. 
Since a time-dependent collision integral is rather difficult to solve, we consider the medium to be large enough such that the formation time is much smaller than the medium length. In this case, the radiation rates are given by the infinite medium limit derived in the AMY approach \cite{Arnold:2000dr}. The full evolution of the phase-space distribution and the details of the implementation of the collision integrals are given in \cite{Mehtar-Tani:2022zwf}.

We will focus on the energy loss of a hard gluon in a static medium of infinite length. The in-medium cascade of the hard gluon leads to a dilute distribution of quarks and gluons compared to the QGP, which we can describe using a linearized fluctuation $\delta f(\p)$ on top of the equilibrium distribution $n_a(\p)$. The full phase-space distribution is then given by
\begin{align}
    f_a(\p) = n_a(\p) + \delta f_a(\p)\;. 
\end{align}
Since the equilibrium distribution $n_a(\p)$ is static, the distribution $ \delta f_a(\p)$ will describe the evolution of the hard partons and the response of the medium. Each elastic scattering with the medium generates recoil partons close to medium scales. The medium parton, which undergoes the scattering, is ``extracted" from the medium. It manifests itself as a negative contribution to the distribution.
\subsection{Evolution of a gluon in a QGP brick}
We consider an initial \emph{gluon} with momentum along the $z$-axis and approximate the initial distribution as a narrow Gaussian centered at $\p = E_0 \hat{e}_{z}$ written as
\begin{align}
    \delta f_g^{\rm in}(p,\theta) &=
    \frac{\exp\left[-\frac{\left(p\cos\theta-E_0\right)^2 + p^2\sin^2\theta}{2\sigma^2 }\right]}{p^3N}, \nonumber\\
    \delta f_{q,\bar{q}}^{\rm in}(p,\theta) &= 0\;,
\end{align}
where $N=\int dp\int d\cos\theta~ \exp\left[-\frac{\left(p\cos\theta-E_0\right)^2 + p^2\sin^2\theta}{2\sigma^2 }\right]$ is a normalization factor. We take the initial energy $E_0=25$~GeV, the QGP temperature $T=0.25$~GeV, the QCD coupling constant to be $g=2$, and the Gaussian width $\sigma = 10^{-3}/\sqrt{2} E_0$.

Typically, jet fragmentation is studied using the Lund plane diagram, which describes jet emissions using its longitudinal momentum $p$ and the inverse of its angle $\theta$ with respect to the primary jet axis \cite{Andersson:1988gp}. In order to understand the chemical composition of the shower, we follow a similar approach to the Lund diagram by studying the distribution of partons in different momenta regions as a function of the polar angle $\theta$ away from the original parton. 

Integrating over a momentum range, we define the following particle number distribution as a function of the polar angle $\theta$ 
\begin{align}
    \frac{d N_a}{d\theta}(\theta) = \sin\theta \int_{p_{\rm min}}^{p_{\rm max}} dp \int \frac{ d\phi}{(2\pi)^2} p^2 \delta f_a(\p)\;.
\end{align}
Figure~\ref{fig:KineticEvol} presents the distribution of gluons in full lines compared with the distribution of quark and antiquarks in dashed lines. The different panels show the angular distribution integrated over two momenta ranges: (top panel) the semi-hard partons with $8T\leq p \leq 20T$ and (bottom panel) soft partons with $p\leq 10T$. 
The evolution at times $t=2$ and 6~fm/$c$ are selected to represent typical times of jet energy loss in the QGP, while $t=10$~fm/$c$ corresponds to a near-equilibrium distributions where most of the hard parton's energy is lost to the medium.

One observes at earlier times ($t=2$ and 6~fm/$c$) that the hard core of the distribution at $\theta\simeq 0$ is composed of more bosons, originating from collinear radiation of the initial gluon. Since the equilibration of bosons proceeds faster than for fermions, one finds slightly more bosons at the low scales of $p\leq 8T$.
However, for the semi-hard partons $8T\leq p\leq 20T$, there is a development of higher number of fermions in a ring with $\theta \gtrsim 0.6$. 
Conversely, at late times ($t=10$~fm/$c$), the fermions dominate over bosons in all momenta ranges and angles as chemical equilibration is reached, leading to the same parton composition as the QGP \cite{Mehtar-Tani:2018zba,Schlichting:2020lef}.

\subsection{Chemical composition at late time}
Throughout the preceding subsection, we followed the evolution of a linearized perturbation on top of a static equilibrium background, which at asymptotically late time completely thermalizes with the medium. 
The asymptotic distribution can be obtained analytically by considering a linear perturbation around the equilibrium distribution $n_a(\p)$ for each species. To achieve this, only the linear terms of the Taylor series in the thermodynamic conjugate of the conserved quantities are kept.

For the kinetic evolution considered, the conserved quantities are energy $E$, momentum $p_z$, and valence number $N_v = N_q - N_{\bar{q}}$, while their conjugate variables are temperature $T$, flow velocity $u_{z}$, and chemical potential $\mu_v = \mu_q - \mu_{\bar{q}}$, respectively. The general equilibrium distribution is 
\begin{align}
    n_a(\p) = \frac{1}{e^{\frac{p\cdot u   - \mu_a }{T}}\mp 1}\;,
\end{align}
where $\mp$ stand for gluons and quarks, respectively. The linear perturbation of the equilibrium distribution is then given by 
\begin{align}
    \delta n_a&(\p) =\\
    &\left. \left[ -\frac{\delta T}{T^2} \partial_T + \delta u_z \partial_{u_z} + \delta \left( \frac{\mu_a}{T}\right) \partial_{\frac{\mu_a}{T}} \right] n_a(\p)\right|_{u_z=\mu_a= 0}\,.\nonumber
\end{align}
After identifying the values of $\delta T$, $\delta u_z$ and $\delta \left( \frac{\mu_a}{T}\right)$ by matching the moments of the distribution with the conserved quantities, one finds\footnote{A detailed derivation is given in App.~C of \cite{Mehtar-Tani:2022zwf}.} \cite{Mehtar-Tani:2022zwf} 
\begin{align}
    \delta n_a(\p) =& E_0 \frac{p}{4T\epsilon(T)} [1+3\cos\theta] n_a(\p)(1\pm n_a(\p))\;,
\end{align}
with the energy density $\epsilon(T) = \left( \frac{\pi^2}{30}\nu_g + \frac{7\pi^2}{120} \nu_q N_f\right)T^4$, where $\nu_g=2(N_c^2-1)$ and $\nu_q = 2N_c$. 
The matching ensures that the energy of the initial parton is recovered by computing the following moment of the distribution, 
\begin{align}
    \int \frac{d^3 \p}{(2\pi)^3} \nu_g \delta n_g (\p) + 2N_f \delta n_q(\p) = E_0\;.
\end{align}

When the momentum of the partons is much larger than the temperature $p\gg T$, the quantum distributions can be approximated by a Boltzmann distribution, leading to a simple relation between the number of fermions and gluons. While the low momentum region $p\ll T$ is dominated by gluons, for large momenta $p\gg T$, the number of fermions is related to the gluon number by  
\begin{align}
    \frac{N_q(\p) + N_{\bar q}(\p)}{N_g(\p)} =
    \frac{2N_f \nu_q \delta n_q(\p)}{\nu_g \delta n_g(\p)} \overset{\p \gg T}{\simeq} \frac{4N_c N_f}{2d_A}\;,
\end{align}
leading to the relation 
\begin{align}
    N_q(\p) + N_{\bar q}(\p) \overset{\p \gg T}{\simeq} \frac{N_f}{C_F} N_g(\p)\;.
\end{align}

In the above section, we demonstrated how the thermalization of a hard gluon in a hot and dense QGP leads to a shower with a chemical composition dominated by quarks and antiquarks, contrary to the case of vacuum fragmentation. Throughout this kinetic theory simulation, we have considered a simple static medium, which ignores important effects of flow. 

Our solution to the Boltzmann equation does not include any event-by-event fluctuations, initial vacuum like shower for partons at large virtualities, or realistic energy loss parameters. In the following section, we will consider realistic simulations where the hard jet shower may not thermalize in the medium. While the medium will remain static, the jet will undergo a stochastic process of multiple emission as it propagates through the medium. We will study cases both with and without vacuum like emissions. We will also consider the systematic effect of varying the energy loss formalism from a single stage to a multiple stage formalism.

\section{Simulations in vacuum and static media}
\label{sec:realistic_simulation}
In the preceding section, we demonstrated that the appearance of a large number of quarks and antiquarks within the vicinity of a jet should be a generic feature of jet modification processes in a deconfined medium. The semi-analytic results in a static medium did not include vacuum like showers~\cite{Caucal:2018dla,Cao:2017qpx,Cao:2017zih} and multi-stage energy loss~\cite{JETSCAPE:2022jer} in a dynamically evolving medium. In this section, somewhat more realistic simulations will be carried out to study the appearance of these charge/baryon rings in the angular structure of jets. 

In the first subsection, we will revisit the calculation of the angular structure of gluons and quarks radiated from a hard \emph{gluon} in vacuum, demonstrating the large excess of gluons at all angles away from the primary parton. Following this, simulations are carried out in a static medium at $T=0.25$~GeV, using the Linear Boltzmann Transport (LBT) event generator, which is somewhat different from the Boltzmann equation based calculations presented in Sec.~\ref{sec:brick}: LBT is a Monte-Carlo event generator and there are at most one scattering per emission, for all emissions.

Finally, simulations with a multi-scale event generator are presented, where the initial high virtuality stage is modeled with the higher twist formalism in the MATTER generator~\cite{Majumder:2013re,Cao:2017qpx} and the lower virtuality stage is modeled using the Hard Thermal Loops formalism~\cite{Arnold:2000dr,Arnold:2002zm,Arnold:2003zc} present in the MARTINI generator~\cite{Schenke:2009gb}, which involves multiple coherent scatterings per emission. Similar to the case of pure LBT, these simulations are also carried out in a static medium.

Compared to the vacuum shower, the pure LBT simulation or the MATTER+MARTINI combination generates an excess of quarks+antiquarks at large angles away from the jet axis at both intermediate and low-$p_T$. The angle at which these appear may vary, based on the parameters of the simulation. In both the LBT and the MATTER+MARTINI simulation, the number of semi-hard quarks and antiquarks exceeds the number of semi-hard gluons by $\tau = 10$~fm/$c$.

\subsection{Simulations in Vacuum}
\label{subsec:vacuum}
We begin by revisiting the partonic angular structure of jets in a vacuum. We consider a hard gluon with $E=25$~GeV that starts with a typical initial maximum virtuality of $Q=E/2$. This implies that the initial virtuality is logarithmically distributed in the range $0 \leq \mu \leq Q$. As in the preceding section, a hard gluon is the shower-initiating parton. This choice removes any contamination of the scattering-generated charge ring from the parent parton, as the gluon has no net charge or baryon number.

\begin{figure}
    \centering
    \includegraphics[width=0.4\textwidth]{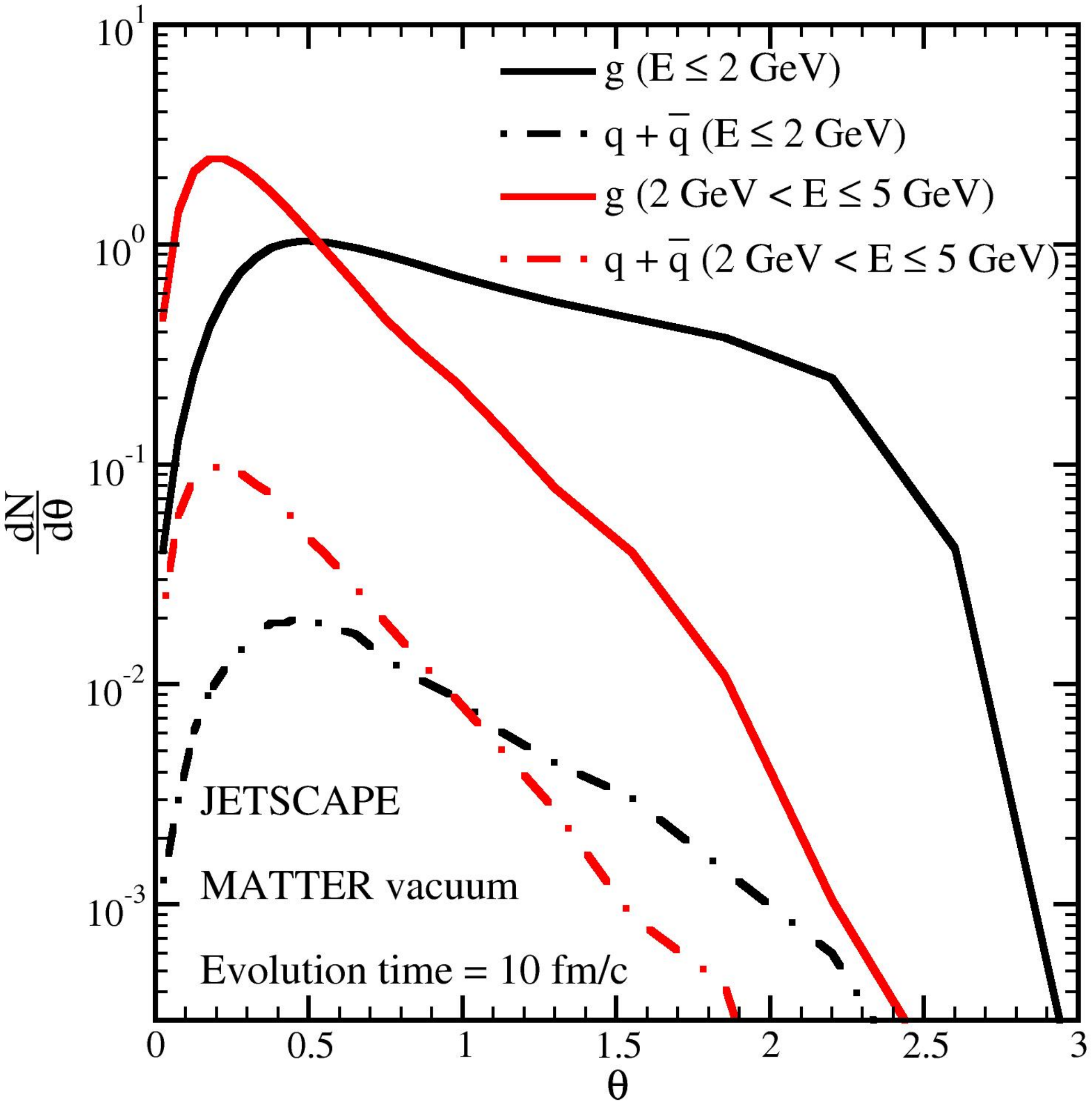}
    \caption{(Color Online) Particle number distribution from a 25~GeV initial gluon in vacuum, as a function of the polar angle, for hard partons with energy $2$~GeV$< E\leq 5$~GeV (red lines), and soft partons with energy $E\leq 2$~GeV (black lines). The gluon distribution is displayed in full-lines, while the sum of quarks and antiquarks is displayed in dashed-lines, at $\tau_{\rm max}=10$~fm/$c$ after the production of the original gluon, which has a maximum virtuality of $E/2=12.5$~GeV.}
    \label{fig:Vacuum}
\end{figure}

The hard parton can undergo successive splits, where both $g \rightarrow gg$ and $g \rightarrow q \bar{q}$ are allowed. The shower development is continued for a time $\tau = 10$~fm/$c$. In Fig.~\ref{fig:Vacuum} we split the final partons at $\tau=10$~fm/$c$ into two groups: The low-energy group with $E < 2$~GeV, and the intermediate energy group with $2$~GeV$< E \leq 5$~GeV. Had the jet been immersed in a medium at a temperature $T=0.25$~GeV, the two energy boundaries would have corresponded to $8T$ and $20T$, similar to the ranges considered in the preceding section.

Given the singular nature of the $g \rightarrow gg$ splits compared to the $g \rightarrow q \bar{q}$ splits, one notes that in both the low and intermediate energy range, the number of gluons far exceeds the number of quarks and antiquarks. As stated in the introduction, vacuum jets are primarily gluonic. Comparison with the plots in Fig.~\ref{fig:KineticEvol}, for the case of pure in-medium evolution, should immediately convince the reader of the striking difference between the jet flavor profile in a medium versus that in a vacuum. 
By 10 fm/$c$, the quark and antiquark population in Fig.~\ref{fig:KineticEvol}, easily exceeds the gluon population in most regions of phase space (green lines in the plot).

\subsection{Simulations in LBT}
\label{subsec:LBT}
In this subsection, we present results for the flavor profile from a semi-realistic Monte-Carlo simulation of a hard gluon propagating through a static medium, held at $T=0.25$~GeV. In this first attempt to reveal the charge/flavor/baryonic profile, simulations will be carried out starting from a 25~GeV gluon within the \emph{pure} LBT model, i.e., using a single stage jet modification scenario. A two-stage simulation is presented in the subsequent subsection. 

In LBT~\cite{Cao:2016gvr,Cao:2017hhk}, the phase space distribution of the jet parton (denoted by $a$) evolves according to the Boltzmann equation as
\begin{equation}
p_a \cdot \partial f_a= E_a (\mathcal{C}_\mathrm{el}+\mathcal{C}_\mathrm{inel}),
\label{eq:Boltzmann}
\end{equation}
in which the collision term on the right hand side incorporates both elastic and inelastic contributions. Based on the collision term, the elastic scattering rate, i.e., the number of elastic scatterings per unit time, as
\begin{align}
\label{eq:rate}
\Gamma_a^\mathrm{el}&(\p_a,T)=\sum_{b,(cd)}\frac{\gamma_b}{2E_a}\int \prod_{i=b,c,d}\frac{d^3 \p_i}{E_i(2\pi)^3} f_b S_2(\hat{s},\hat{t},\hat{u})\nonumber\\
&\times (2\pi)^4\delta^{(4)}(p_a+p_b-p_c-p_d)|\mathcal{M}_{ab\rightarrow cd}|^2,
\end{align}
in which the summation is over all possible $ab\rightarrow cd$ channels, $\gamma_b$ represents the color-spin degrees of freedom of the thermal partons inside the QGP and $f_b$ is their distribution function. In LBT, a function $S_2(\hat{s},\hat{t},\hat{u})=\theta(\hat{s}\ge 2\mu_\mathrm{D}^2)\theta(-\hat{s}+\mu^2_\mathrm{D}\le \hat{t} \le -\mu_\mathrm{D}^2)$ is introduced~\cite{Auvinen:2009qm} to regulate the collinear divergence in the leading-order (LO) scattering matrices $\mathcal{M}_{ab\rightarrow cd}$, where $\hat{s}$, $\hat{t}$ and $\hat{u}$ are the Mandelstam variables and $\mu_\mathrm{D}^2=g^2_sT^2(N_c+N_f/2)/3$ is the Debye screening mass with $g^2_s=4\pi \alpha_s$ being the strong coupling constant and $T$ being the medium temperature.

The inelastic scattering rate can be related to the average number of medium-induced gluons per unit time as
\begin{equation}
 \label{eq:gluonnumber}
 \Gamma_a^\mathrm{inel} (E_a,T,t) = \int dxdk_\perp^2 \frac{dN_g^a}{dx dk_\perp^2 dt},
\end{equation}
with the gluon spectrum taken from the higher-twist energy loss calculation~\cite{Wang:2001ifa,Zhang:2003wk,Majumder:2009ge},
\begin{equation}
\label{eq:gluondistribution}
\frac{dN_g^a}{dx dk_\perp^2 dt}=\frac{2C_A\alpha_\mathrm{s} P^\mathrm{vac}_a(x)}{\pi C_2(a) k_\perp^4}\,\hat{q}_a\, {\sin}^2\left(\frac{t-t_i}{2\tau_f}\right).
\end{equation}
Here, $x$ and $k_\perp$ are the fractional energy and transverse momentum of the emitted gluon relative to its parent parton, $P^\mathrm{vac}_a(x)$ is the vacuum splitting function of the jet parton with its color factor $C_2(a)$ included, $\hat{q}_a$ is the jet quenching parameter that encodes the medium information and is evaluated
according to the transverse momentum broadening square per unit time in elastic scatterings, $t_i$ denotes the production time of parton $a$, and $\tau_f={2E_a x(1-x)}/k_\perp^2$ is the formation time of the emitted gluon. In this section, we set the coupling constant as $\alpha_{s}=0.3$, which directly controls the interaction strength in elastic scatterings, and affects the rate of medium-induced gluon emission through $\hat{q}_a$.

\begin{figure}[htbp!]
    \centering
    \includegraphics[width=0.4\textwidth]{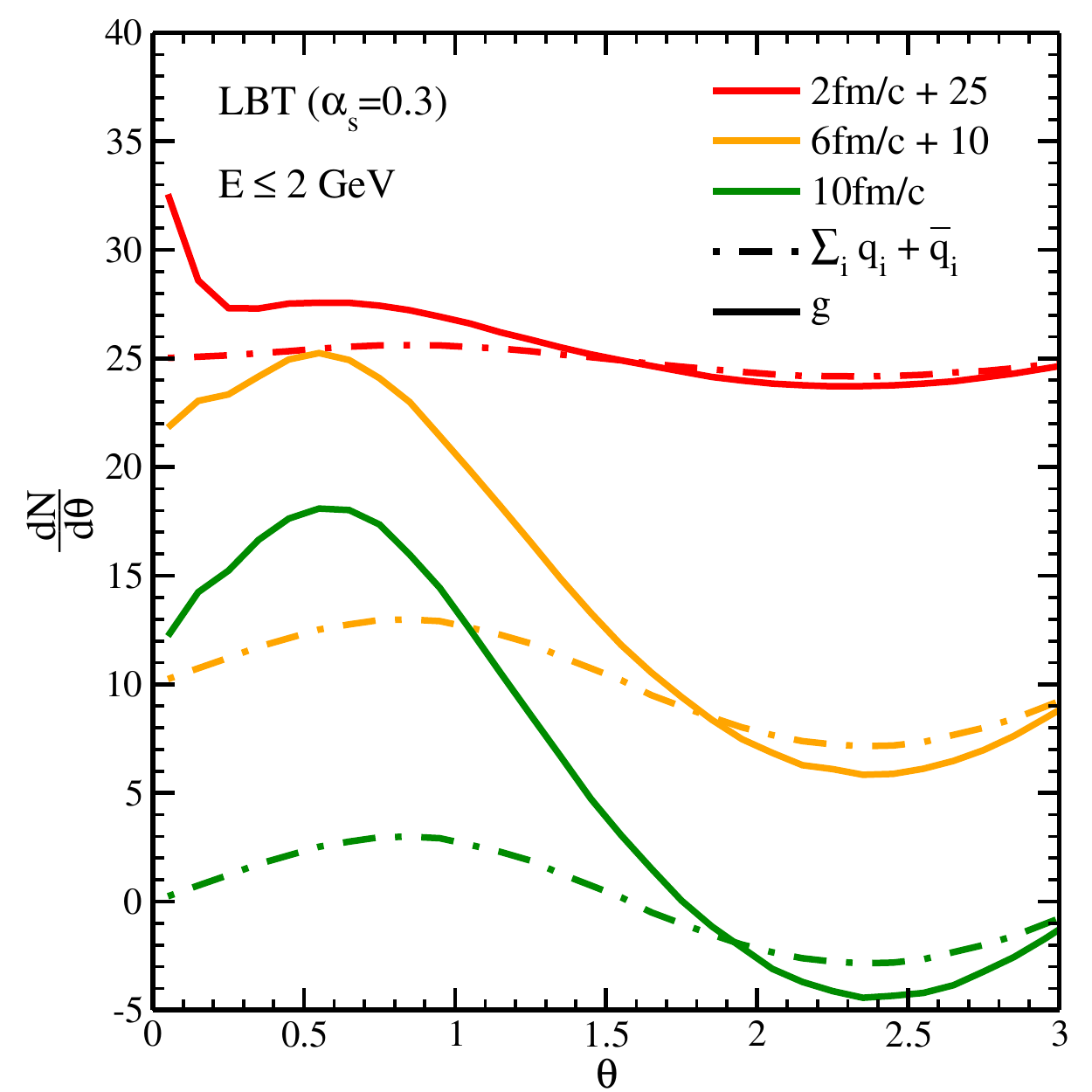}
    \caption{(Color Online) Particle number distribution from LBT simulation starting from a $E_{\rm in}=25$~GeV gluon and evolving in a static medium at $T=0.25$~GeV, as a function of the polar angle, for soft partons with energy $E\leq 2$~GeV. The gluon distribution is displayed in full-lines, while the sum of quarks and antiquarks is displayed in dashed-lines, at three evolution times of 2, 6 and 10~fm/$c$.}
    \label{fig:LBT_low_pT}
\end{figure}

In the LBT simulation\footnote{In this particular simulation, the rate for gluons to convert to quarks or antiquarks has been corrected by an additional factor of $N_f=3$, which is missing in all prior versions. However, this factor should not change previous results which focused on studying energy loss.}, we track not only the jet partons and their emitted gluons, but also the thermal partons being scattered out of the QGP background by jets. The latter are known as ``recoil" partons. When these ``recoil" partons are produced, energy-momentum depletion occurs inside the original QGP medium. These are treated as particle holes, or ``negative" partons, and also fully tracked in LBT in order to guarantee the energy-momentum conservation of the whole system of jet partons and the QGP. Recoil and ``negative" partons constitute the ``jet-induced medium excitation", or ``medium response to jet propagation", which have been shown to be crucial for understanding jet observables, including their nuclear modification factor and anisotropic flow coefficients~\cite{He:2018xjv,He:2022evt}.

\begin{figure}[htbp!]
    \centering
    \includegraphics[width=0.4\textwidth]{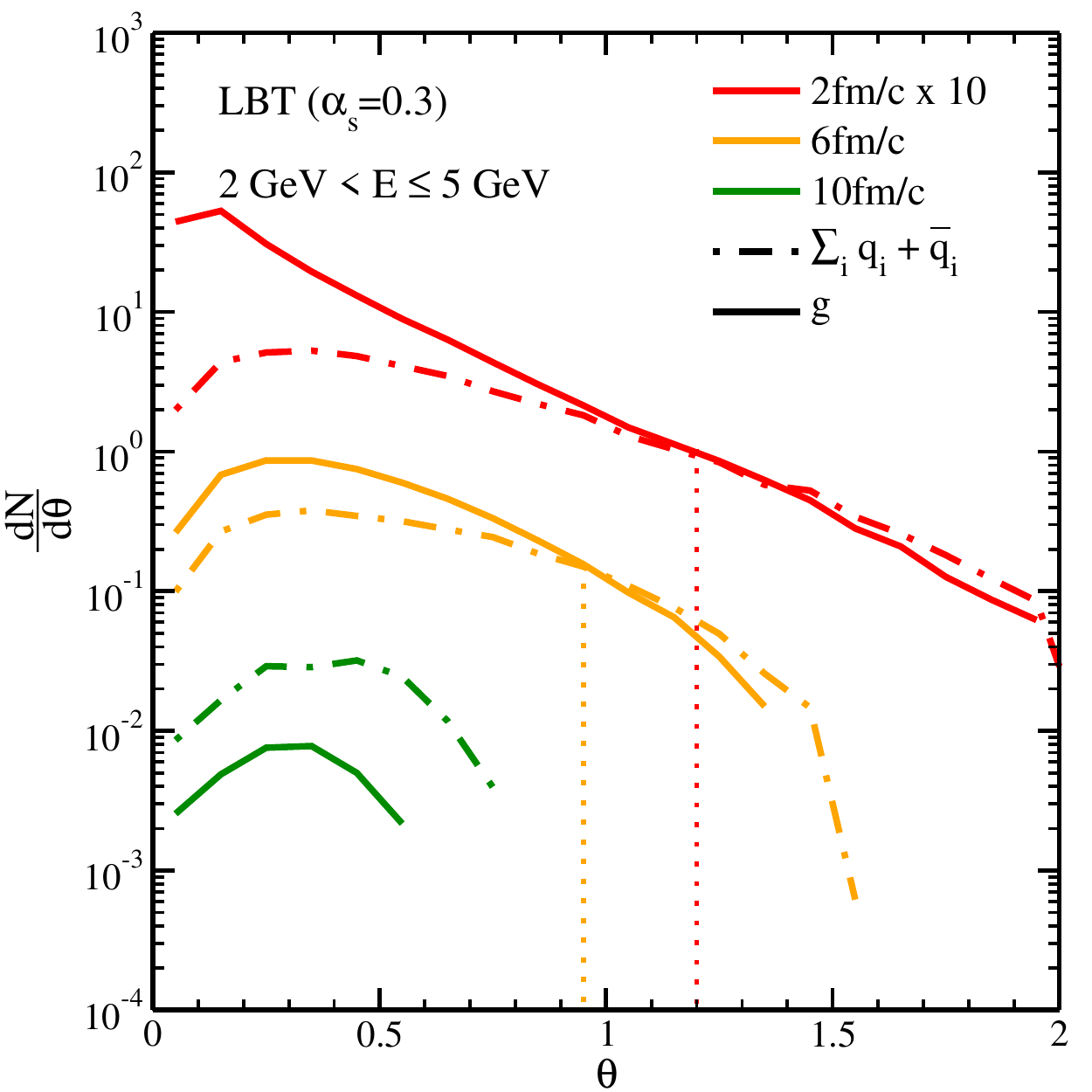}
    \caption{(Color Online) Same as simulations in Fig.~\ref{fig:LBT_low_pT}, except for partons at intermediate energy $2< E \leq$ 5~GeV. The dotted lines show the angles quark distribution starts to exceed the gluon distribution for the evolution times 2 and 6~fm/$c$.}
    \label{fig:LBT_intermediate_pT}
\end{figure}

Using this LBT model, we calculate the angular distribution of partons that start from a single gluon with 25~GeV energy and evolve through a static medium at $T=0.25$~GeV. Results for partons at low energy ($E\le 2$~GeV) and intermediate energy ($2<E\le 5$~GeV) are presented separately in Figs.~\ref{fig:LBT_low_pT} and~\ref{fig:LBT_intermediate_pT} respectively. In each figure, we compare the distributions of quarks + antiquarks and gluons at three different evolution times. 

At intermediate energy, one can clearly observe an excess of quarks (together with antiquarks) over gluons at larger angles ($\theta \gtrsim 0.9$) with respect to the jet direction (the momentum direction of the initial gluon here), for evolution times up to 6~fm/$c$. At later times, the fermion excess at intermediate momenta manifests at all angles. This excess becomes more prominent as the evolution time increases, indicating a flavor change from gluons to quarks during jet-medium interactions. Note that within the LBT calculation, the distributions of ``negative" partons have been subtracted from those of regular partons. For this reason, one can see the negative distribution of low energy partons (Fig.~\ref{fig:LBT_low_pT}) at large angle, which is known as the energy depletion, or the diffusion wake, in the opposite direction of jet propagation.

While the distribution of soft partons with $E<2$~GeV (Fig.~\ref{fig:LBT_low_pT}) produced in the LBT simulation are rather different from the soft parton distributions in the prior Boltzmann simulation (lower panel of Fig.~\ref{fig:KineticEvol}), the semi-hard distributions (in Fig.~\ref{fig:LBT_intermediate_pT}) are in qualitative agreement with those in the upper panel of Fig.~\ref{fig:KineticEvol}. However, there is almost a factor of 2 difference in the overall normalization of the plots for distributions with $t\gtrsim 6$~fm/$c$. Also, the detailed positions at which quark spectra cross gluon spectra, indicated by the vertical dashed lines are quantitatively different due to different model implementations. In spite of these differences, in both cases, the semi-hard quark (and antiquark) distribution begins to surpass the semi-hard gluon distribution at $\theta \gtrsim 0.2$ for the Boltzmann simulation, and at $\theta \gtrsim 0.9$ for the LBT simulation after 6~fm/$c$, and completely dominates by 10~fm/$c$.
 
\subsection{Simulations in MATTER+MARTINI}

Currently, multi-stage jet modification simulators~\cite{Kumar:2019bvr,Putschke:2019yrg,JETSCAPE:2022jer} have shown remarkable success in simultaneously describing a host of jet-based observables. In these simulations, the medium generated scale $Q^2_{\rm med} = \sqrt{2 E \hat{q}}$, where $E$ is the energy of a parton undergoing energy loss, plays a crucial role~\cite{Majumder:2010qh, Cao:2017zih}. Partons whose virtuality is above this scale undergo mostly vacuum like splitting, with a perturbative correction to the splitting kernel from medium induced radiation. As a result, most emissions are vacuum-like with a few interfering medium induced emissions~\cite{Majumder:2009zu, Kumar:2019uvu,Cao:2021rpv}. In-medium scatterings are accounted for using the scattering kernels described in Sec.~\ref{sec:rates}. As those rates are obtained assuming the incoming and outgoing partons are on-shell, the virtuality is temporarily removed from the $p^0$ component of the four-momentum of incoming and outgoing partons when computing the scattering rates. Once the four-momenta of all partons participating in the scattering is determined, the virtuality is restored within the energy of all incoming and outgoing partons, thus preserving its value. Partons with a virtuality at or below this scale undergo multiple scattering in the process of almost every emission, with purely vacuum like emission almost absent~\cite{Baier:1994bd,Baier:1996sk}.

\begin{figure}
    \centering
    \includegraphics[width=0.4\textwidth]{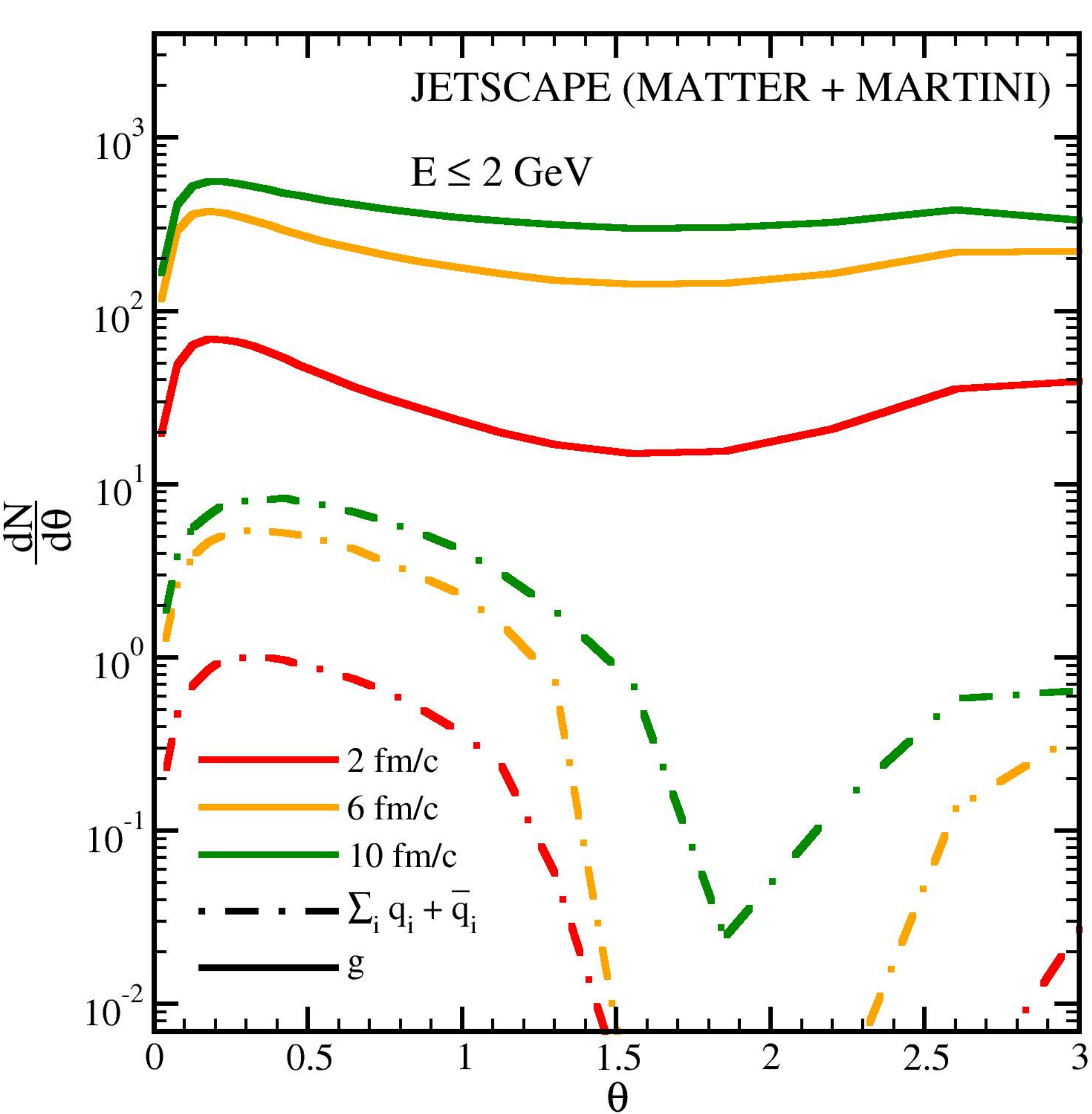}
    \caption{(Color Online) Particle number distribution from a $E_{\rm in}=25$~GeV gluon in a static medium at $T=0.25$~GeV, as a function of the polar angle, for soft partons with energy $E\leq 2$~GeV. The gluon distribution is displayed in full-lines, while the sum of quarks and antiquarks is displayed in dashed-lines, at three evolution times of 2, 6 and 10 fm/$c$ after the production of the original gluon, which has a maximum virtuality of $E/2=12.5$~GeV.}
    \label{fig:MATTERMARTINI_low_pT}
\end{figure}

Simulations in this subsection are carried out using the JETSCAPE framework~\cite{Putschke:2019yrg}, using the version of MATTER and MARTINI simulation modules therein. We consider, once again the case of a single hard \emph{gluon} with an energy of 25~GeV propagating in a static medium held at $T=0.25$~GeV. The hard jet starts with an initial maximal virtuality $Q=E/2$ as in the case of the vacuum simulation in Sec.~\ref{subsec:vacuum}. The emissions from partons with a virtuality $Q > Q_{\rm med}$ are simulated using the MATTER generator. As partons undergo more splits in MATTER, their virtuality drops. Once a parton reaches the $Q_{\rm med}$, it transitions to the MARTINI generator. The virtuality of the partons is maintained by scattering in the medium while in the MARTINI stage. 

As the parton emerges from the medium, $\hat{q}$ will drop to zero, and the virtuality of the parton will once again exceed the scale $Q_{\rm med} \rightarrow 0$, and the parton will transition back to MATTER again. Partons that escape the medium will continue to endure vacuum like splits until each of their virtualities reaches $Q_0 = 1$~GeV. Beyond this, partons will free stream until the end of the simulation, set at $\tau_{\rm max}$.

While the MATTER generator involves at most one scattering per emission from a parton, MARTINI allows for multiple scatterings over the course of a single emission, and as a result there is a greater probability to convert a boson into a fermion (and vice versa), especially for the longer-lived (softer) partons in the MARTINI phase. However, since some portion of the jet will definitely be in the MATTER stage, fewer conversions are expected within a multi-stage parton energy loss simulation compared to a pure MARTINI simulation.

\begin{figure}
    \centering
    \includegraphics[width=0.4\textwidth]{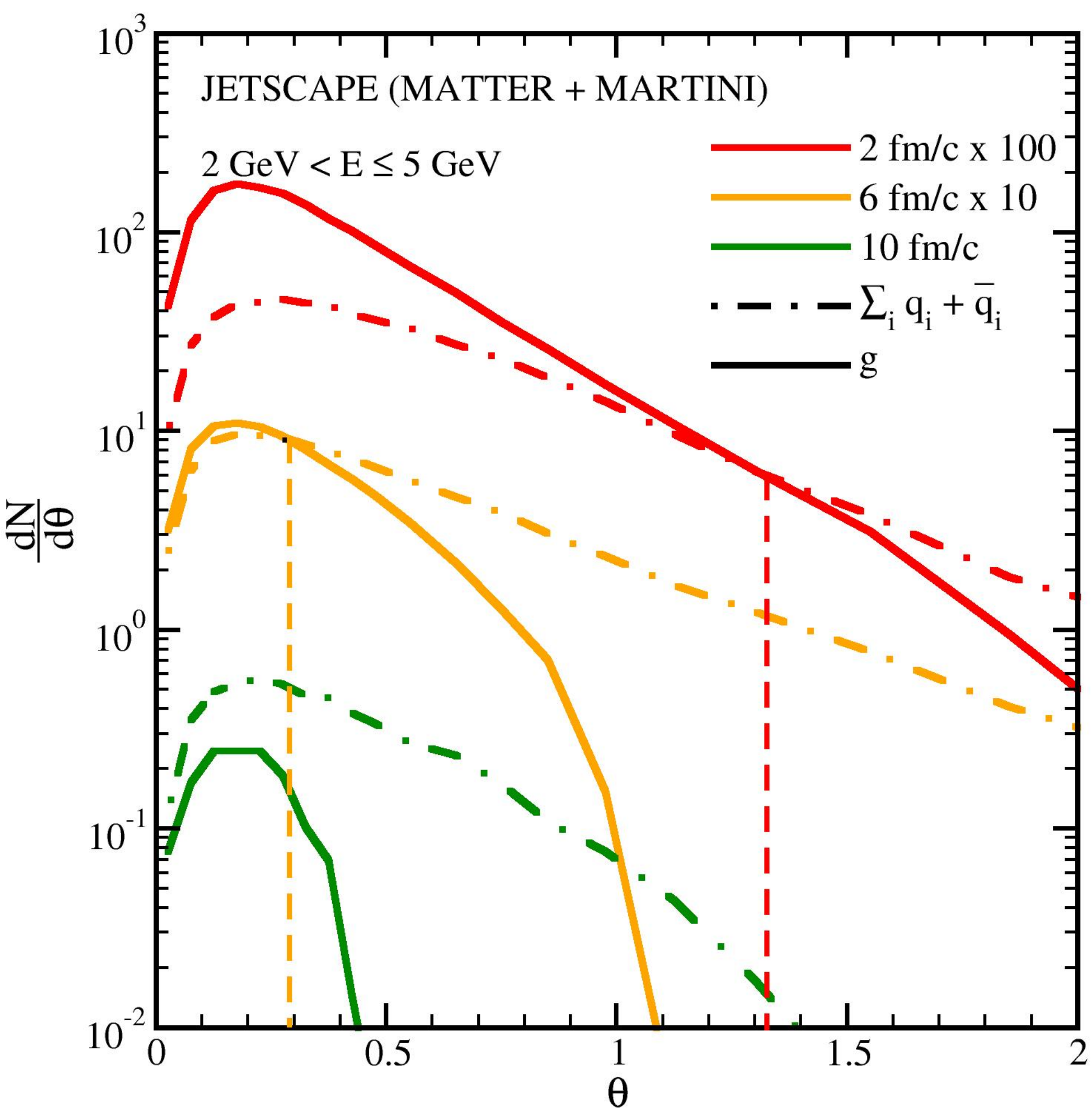}
    \caption{(Color Online) Same as simulations in Fig.~\ref{fig:MATTERMARTINI_low_pT}, except for partons at intermediate $p_T$, with 2 $< E \leq$ 5~GeV. The dashed lines show the angles quark distribution starts to exceed the gluon distribution for the evolution times 2 and 6~fm/$c$. }
    \label{fig:MATTERMARTINI_intermediate_pT}
\end{figure}

In Figs.~\ref{fig:MATTERMARTINI_low_pT} and~\ref{fig:MATTERMARTINI_intermediate_pT}, the yield of semi-hard partons and soft partons, respectively, has been plotted for 3 different values of $\tau_{\rm max}$. These partons are all part of the profile of the jet that starts as a single gluon with an energy of 25~GeV (and virtuality $Q=E/2$). Similar to the case of the solution of the Boltzmann equation in Sec.~\ref{sec:brick}, as well as for the case of LBT in the preceding subsection, we find more quarks and antiquarks in proportion to the gluons, compared to the case in vacuum. In this case, the angle at which the quark and antiquark number exceeds the gluon number at 6 fm/$c$ ($\theta \gtrsim 0.3$), is smaller than in the case of LBT ($\theta \gtrsim 0.9$), and slightly larger than the angle in the Boltzmann simulation ($\theta \gtrsim 0.2$). This is due to greater number of scatterings in MARTINI compared to LBT, and lack of a vacuum like stage in the Boltzmann simulations compared to the MATTER+MARTINI simulations.

The focus of this article is to compare the temporally rising quark (and antiquark) distribution, within and surrounding a jet, with the falling gluon distribution, in the same region of angular space, at intermediate momentum (2~GeV $\lesssim E \lesssim$ 5~GeV). While, the low momentum region is not our focus, we report on it for all three cases of the Boltzmann simulation, LBT and MATTER+MARTINI simulations. In all three cases, the low momentum region around the jet in these simulations is quite different. In the specific case of the MATTER+MARTINI simulation, the number of gluons always remains larger than the number of quarks and anti-quarks. Also, both quark and gluon curves show a dip around $\theta \gtrsim \pi/2$ from the direction of the leading parton. This is primarily due to the subtraction of holes. In the case of the LBT simulations, this region is actually negative (see Fig.~\ref{fig:LBT_low_pT}). In the case of MARTINI, jets can emit partons down to vanishingly soft momentum, which is enhanced for the case of gluons. As a result, the soft gluon emissions completely cover up the negative portion that arises due to subtraction of the holes. The soft quark emission rates are much smaller, and thus can only overcome the negative contribution of hole subtraction at times larger than 10~fm/$c$. 

In this and the preceding section we have explored jets in a medium, albeit static, from a variety of formalisms, which have varying amounts of interaction between the jet and the medium. In all cases, we observe a large excess of the fermion number correlated with the jet (compared to a vacuum shower) at angles greater than 0.2 (Boltzmann simulation)  to 0.9 radian (LBT) away from the original jet-axis. The three simulations are quite different and yield very different distributions for partons with $E\lesssim 2$~GeV. However, these differences at low momentum make the qualitative similarities at intermediate momentum a more rigorous prediction of the gluon versus quark and antiquark number. 

The reader will have noted that all our calculations are entirely partonic. Will this charge enhancement survive hadronization in a ring form? Can it be observed in experimental data? The answer to these questions is so far unsettled. Indeed, most of the fermion excess is at low and intermediate $p_T$ where there are no good hadronization mechanisms that can conserve charge and baryon number, either event-by-event, or within angular/rapidity ranges. Cooper-Frye Hadronization~\cite{Cooper:1974mv} is carried out on distributions. The presence of the large number of co-moving quark and antiquarks will lead to very low mass strings if Lund hadronization were applied, leading to a breakdown of that methodology~\cite{Andersson:1997xwk}.
 In the subsequent and penultimate section, we will explore other observables that may be correlated with this enhancement in baryon/charge number, which may already have been observed. 
\section{Jet modification and the baryon enhancement}
\label{baryon_enhancement}

In the preceding sections, we have argued that jets modified in a dense plasma have a strikingly different flavor profile compared to jets in vacuum. Jets in vacuum that begin with either a hard quark or gluon, tend to radiate a large number of gluons, compared to quarks or antiquarks. As shown in Fig.~\ref{fig:Vacuum}, the number of soft gluons ($E < 2$~GeV) exceeds the number of quarks and antiquarks by two orders of magnitude, while the number of intermediate energy gluons (with $2$~GeV$ < E < 5$~GeV) is an order of magnitude larger than quarks and antiquarks of similar energy. This flavor mixture is dramatically changed for the case of jets modified in the medium,  where the quark and antiquark number becomes comparable to the gluon number. All our estimates are based on a jet that starts as a gluon with $E=25$~GeV, 

Three different simulations carried out in the preceding section indicate that the increase in fermion content of the jet is the most dramatic modification of the jet in a dense medium, the fractional change in flavor far exceeds the fraction of energy lost by the jet on passage through the medium. To be clear, a change in the momentum profile of the jet is not expected as a result of this enhancement of fermionic content: There is no excess or depletion in the amount of energy loss of the jet caused by this change in the flavor profile of soft and semi-hard partons. However, one would expect the flavor or baryon number profile of the jet to be modified, especially in the semi-hard region.

Currently, there is no reliable hadronization mechanism that can be used to test this hypothesis, on a triggered jet. However, we can look for such an enhancement in the yield of hadrons at intermediate $p_T$. In the absence of a reliable hadronization mechanism, we propose the somewhat tenuous equivalence in the ratios:
\begin{align}
   \frac{\frac{ d^3 N^{B + \bar{B}}_{AA}(b_{min}, b_{max}) }{d^2 p_T dy} }{\langle N_{bin} \rangle_{(b_{min}, b_{max})} \frac{d^3 N^{B + \bar{B}}_{pp} }{d^2p_T dy } } 
  &\sim \frac{ \frac{ d^3N^{q + \bar{q}}_{AA}  (b_{min}, b_{max})}{d^2p_T dy}  }{  \langle N_{bin} \rangle_{(b_{min}, b_{max})} \frac{d^3 N^{q + \bar{q}}_{pp} }{d^2p_T dy} } \nonumber \\
  \implies   R^{B+\bar{B}}_{AA} &\sim   R^{q+\bar{q}}_{AA} .
\end{align}
In the above equation, we are proposing the $R_{AA}$ for baryons and anti-baryons as an approximation to the $R_{AA}$ for quarks and antiquarks. This equality will no doubt receive corrections from hadronization. We will study this ratio in the intermediate $p_T$ region. The goal is to see the proximity of the two ratios, to place constraints on the possible hadronization mechanisms~\cite{Fries:2003vb, Molnar:2003ff, Hwa:2002tu, Hwa:2004ng, Greco:2003xt} in this region. 

Simulations of this ratio are carried out using the LBT model~\cite{He:2015pra} for energy loss (see Subsec.~\ref{subsec:LBT} for more details). Calculations are carried out on a realistic fluid dynamical medium~\cite{Shen:2014vra}. The initial state and evolution of the fluid have been parameterized by comparison with the yield and azimuthal anisotropy of soft hadrons. The initial hard spectrum of partons has been calculated using LO pQCD, with requisite K-factors~\cite{Field:1989uq}. We present results for the $R_{AA}$ of quarks and antiquarks at 0-20$\%$  central collisions at RHIC ($\sqrt{s_{\rm NN}} = 0.2$~GeV) and 0-5$\%$ collisions at LHC ($\sqrt{s_{\rm NN}} =2.76$~GeV) energies. 

\begin{figure}
    \centering
    \includegraphics[width=0.4\textwidth]{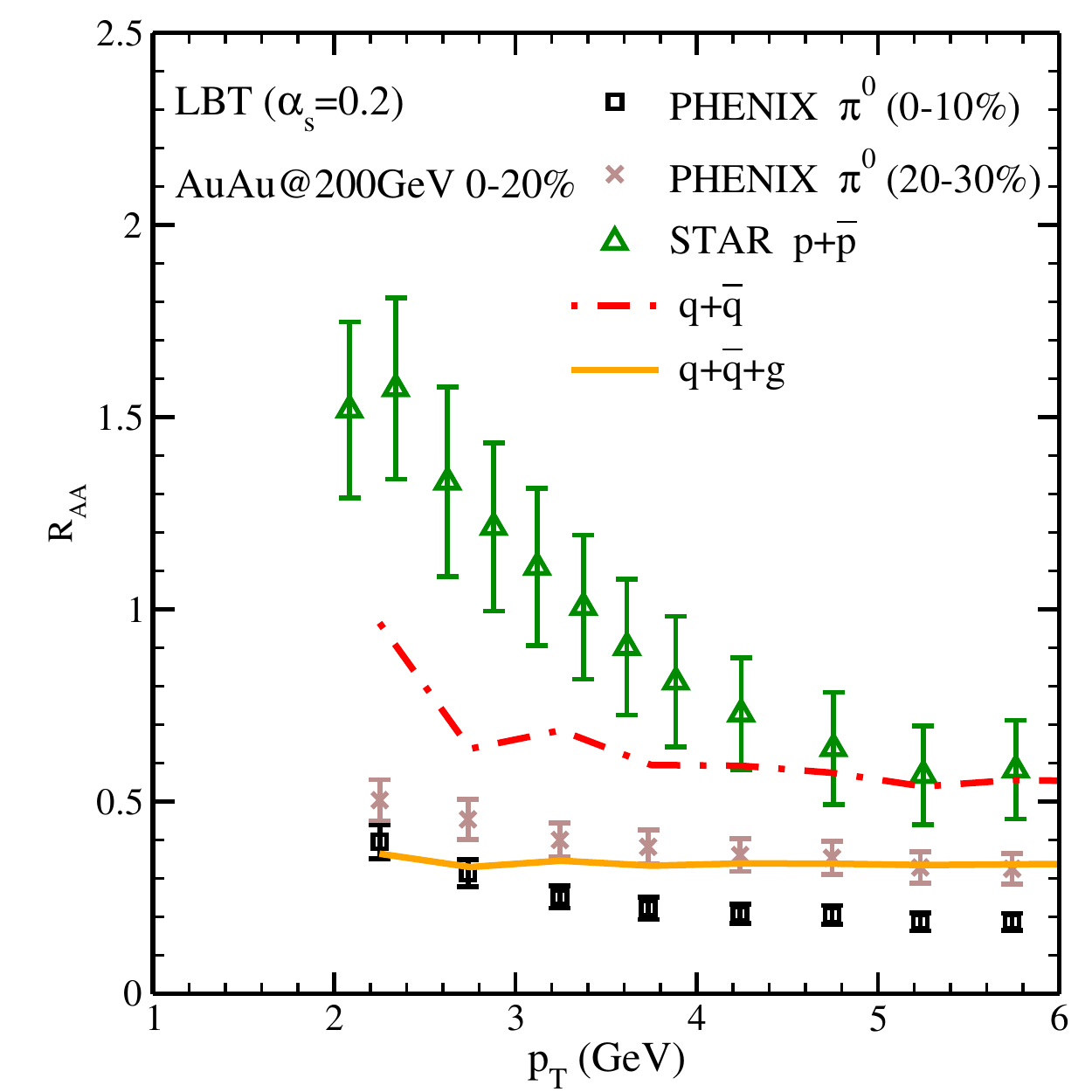}
    \caption{(Color online) The nuclear modification factor for quarks plus antiquarks (red dot-dashed line), and all partons (orange solid line) correlated with hard scattering. The partons included were created in the modified shower from the jet, either via a split from another parton, or via the recoil process. No partons from the fluid, except those in recoil are included. Results are compared with the $R_{AA}$ for $p+\bar{p}$ as an approximate substitute for the baryon plus antibaryon ratio. No hadronization is included in the theoretical calculation. Experimental results taken from Ref.~\cite{PHENIX:2008saf,STAR:2009mwg}.}
    \label{fig:q_qbar_RHIC}
\end{figure}

\begin{figure}
    \centering
    \includegraphics[width=0.4\textwidth]{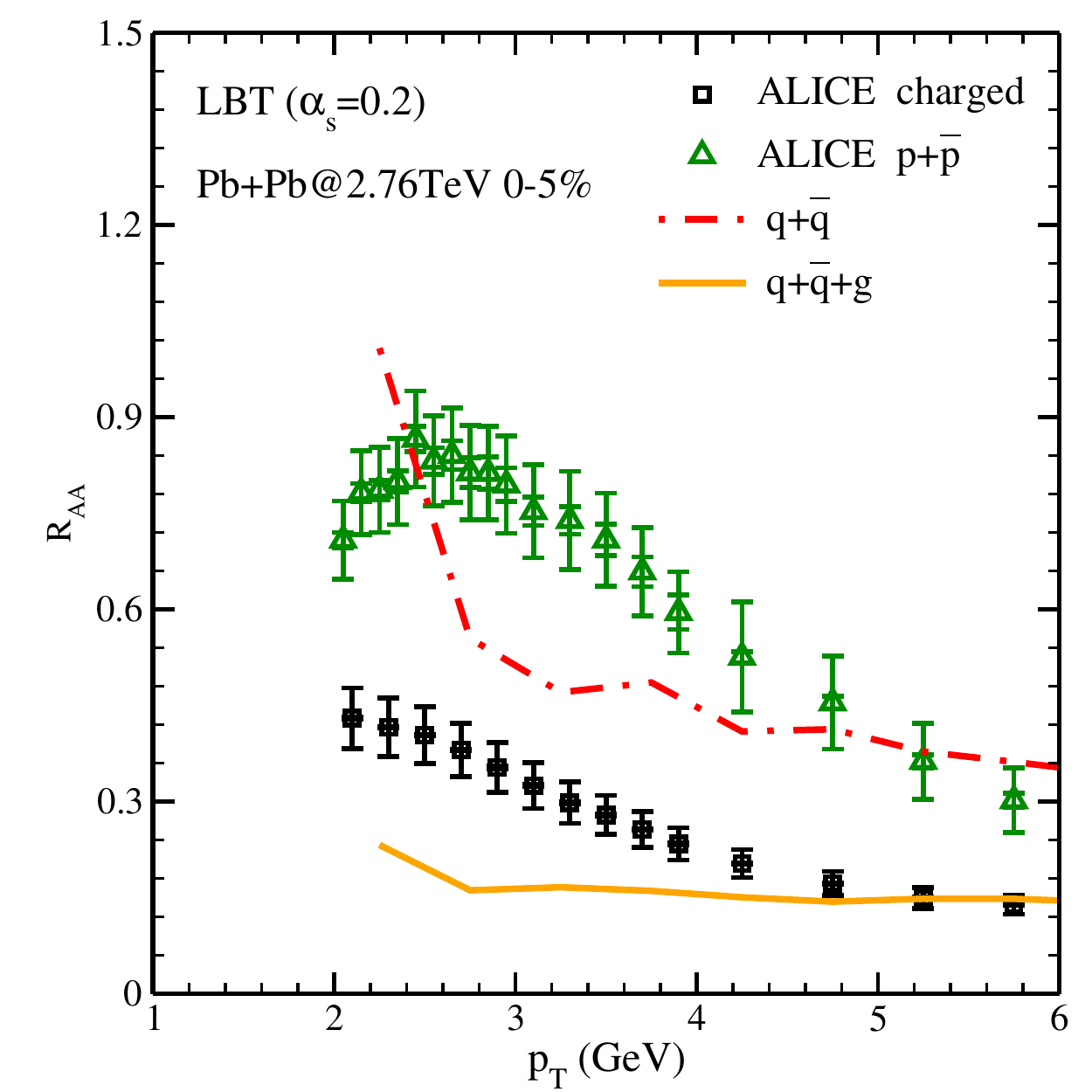}
    \caption{(Color online) Same as Fig.~\ref{fig:q_qbar_RHIC}, for collisions at $\sqrt{s_{\rm NN}}=2.76$~TeV, at the LHC. Experimental results taken from Ref. \cite{ALICE:2014juv,ALICE:2019hno}. }
    \label{fig:q_qbar_LHC}
\end{figure}

In Figs.~\ref{fig:q_qbar_RHIC} and~\ref{fig:q_qbar_LHC}, we plot the nuclear modification factor for partons correlated with hard scattering (solid line in both figures). We include only partons that are created in the split of another parton from the jet showers or created in the recoil process. We also plot the $R_{AA}$ of $q + \bar{q}$ correlated with jets. These are compared with the $R_{AA}$ for proton + antiprotons as a substitute for the baryon and antibaryon $R_{AA}$.

At both RHIC and LHC energies, we note that the $R_{AA}$ for quarks and antiquarks shows a rise at lower $p_T$ that is similar to the rise of the $R_{AA}$ for protons and antiprotons. However, the rise takes place at a lower $p_T$ than the experimental data. Also, the magnitude of the excess at RHIC is less than the data. Thus, the fermion excess from jets may not be sufficient to explain the baryon excess seen at intermediate $p_T$ at RHIC and LHC. However, it may be a part of a multi-aspect solution to this problem. 
It may provide further constraints on recombination models, which have so far been tuned to data without the fermion enhancement. 

An alternative way to view the results in Figs.~\ref{fig:q_qbar_RHIC} and~\ref{fig:q_qbar_LHC} is that for this particular signal, the fermion fraction of a jet, LBT is not an accurate simulator. As pointed out in the preceding section, the MARTINI generator produces many more fermions than LBT as each hard parton in MARTINI undergoes much more scattering compared to LBT. Full simulations in MARTINI on a hydrodynamic background are very computationally demanding and will not be presented in this first effort.

Yet another possibility is that our assumption of completely perturbative interaction between the jet partons and the medium is not accurate and non-perturbative matrix elements will have to be measured and incorporated within these full simulations to reproduce the baryon-antibaryon $R_{AA}$. These non-perturbative matrix elements were discussed in Sec.~\ref{survey} and involve quark matrix elements in the medium. These have so far not been calculated on the lattice or measured in experiment. 
\\

\section{Summary and outlook}
\label{sec:summary_outlook}

The modification of hard jets in a dense QCD medium has traditionally been understood in terms of an increase in the number of gluons radiated from the originating hard parton, followed by a redistribution of the radiated partons towards larger angles away from the jet axis. In this paper, we explored another sizable effect, an order of magnitude increase in the fermion content at intermediate momenta correlated with the jet. 

The origin of this fermion excess, which manifests as an increase in the baryon (and antibaryon) and  charge (and anticharge) distributions at intermediate $p_T \gtrsim 8 T$ ($T$ is the temperature), at angles greater than 0.2-0.5 away from the jet axis, lies predominantly in the recoil mechanism. The rate of a semi-hard gluon scattering off a thermal quark or gluon and converting into a semi-hard quark or antiquark is several times larger than the rate of a semihard quark or antiquark converting into a semi-hard gluon. While these conversion rates are much smaller than the rates of typical scattering, which do not lead to flavor conversion, it is still large enough that a majority of gluons experience at least one such scattering within media with sizes between 5-10~fm/$c$, at average temperatures of approximately 0.25~GeV (values representative of collisions at RHIC and LHC energies).

All these conversion processes involve the exchange of a quark or antiquark with the medium. This feature differentiates these processes from typical scattering in the medium, mediated by gluon exchange, which is typically encapsulated within the well known transport coefficients such as $\hat{q}$ and $\hat{e}$. The quark exchange, manifest in these processes, requires the incorporation of new transport coefficients within the discussion of jet quenching, which will yield insight into the fermion fraction of the underlying evolving medium. 

While partons at an intermediate $p_T \gtrsim 8T$, (typically 2-5~GeV) tend to have a considerable non-perturbative portion in their interaction with the medium, we have carried out this first exploration assuming an entirely perturbative approximation. In spite of this, we find that the fermion fraction correlated with a jet increases ten-fold for jets quenched in a medium, compared to those in vacuum. We considered jets with energies $E\gtrsim 25$~GeV traversing 5-10~fm/$c$ in a medium held at $T\simeq 0.25$~GeV. These are typical distances and temperatures encountered by jets as they traverse media at RHIC and LHC. 

The size of this effect is striking, the number of semi-hard fermions in jets increases by at least an order of magnitude. We have checked this enhancement through three separate model calculations: a semi-analytic solution of the Boltzmann equation, a single stage LBT model and a multi-stage MATTER+MARTINI approach. All three approaches showed similar levels of enhancement of the fermion distribution in jets, compared to the gluon distribution. This has not been pointed out before. 

While the fermion enhancement does not change the energy profile of the jet, it should strongly affect the event-by-event fluctuations of conserved charges such as baryon number and electric charge within the jet. With more charges and anti-charges produced, many of these will be clustered within a jet and many will not; this should lead to larger event-by-event fluctuations of baryon number and charge within a jet quenched in a dense medium, compared to one in vacuum. Of course, the conclusions in this paper will be affected by hadronization, which will introduce its own fluctuations of conserved charges. It is also possible that the large fermionic content introduces an additional source of jet energy loss in the hadronization process. We leave this topic and more realistic simulations of jets in dynamical media for a future effort.

\acknowledgements
The authors thank S.~Shi, C.~Gale and S.~Jeon for help with the MARTINI code. The MATTER+MARTINI simulations were run using the public JETSCAPE code base. A. M. thanks A.~Kumar and J.~Weber for discussions that formed the genesis of this effort. C.~S., I.~S., G.~V. and A.~M. were supported by the US Department of Energy under grant number DE-SC0013460, and by the National Science Foundation under grant numbers ACI-1550300 and OAC-2004571 within the framework of the JETSCAPE collaboration. G.~V. was also supported by Natural Sciences and Engineering Research Council (NSERC) of Canada. W.~J.~X. and S.~C. were supported by the National Natural Science Foundation of China (NSFC) under grant numbers 12175122 and 2021-867.
The kinetic simulation used resources of the National Energy Research Scientific Computing Center, a DOE Office of Science User Facility supported by the Office of Science of the U.S. Department of Energy under Contract No. DE-AC02-05CH11231.

\bibliography{refs}

\end{document}